\documentclass[usenatbib,usegraphicx,fleqn,12pt]{mn2e}
\usepackage{amsmath,amssymb,bm,color,multirow,multicol}
\usepackage{charter}   

\pdfoutput=1
\newcommand{\e}[1]{\times 10^{#1}}

\usepackage{ulem}

\title[Simulations of CTTS V2129 Oph]{Global 3D Simulations of Disc Accretion onto the classical T Tauri Star
V2129 Oph}

\author[M. M. Romanova et al.]
{M. M. Romanova,$^1$\thanks{e-mail:romanova@astro.cornell.edu}, M. Long,$^2$\thanks{e-mail:longm@illinois.edu},
F. K. Lamb$^{2,3}$, A. K. Kulkarni $^1$ and J.-F. Donati $^4$\\
$^1$ Department of Astronomy, Cornell University, Ithaca, NY 14853-6801, USA\\
$^2$ Center for Theoretical Astrophysics, Department of Physics, University of Illinois at Urbana-Champaign, Urbana, IL 61801-3080\\
$^3$ Also Department of Astronomy, University of Illinois at Urbana-Champaign, Urbana, IL 61801-3080\\
$^4$ LATT-UMR 5572, CNRS\& Univ. P. Sabatier, 14 Av. E. Belin, F-31400 Toulouse, France}

\begin{document}



\maketitle


\begin{abstract}

The magnetic field of the classical T Tauri star V2129 Oph can be
modeled approximately by superposing slightly tilted dipole and
octupole moments, with polar magnetic field strengths of 0.35~kG
and 1.2~kG respectively (\citealt{dona07}, hereafter D07). Here we
construct a numerical model of V2129~Oph incorporating this result
and simulate accretion onto the star using a three-dimensional
magnetohydrodynamic code. Simulations show that the disk is
truncated by the dipole component and matter flows towards the
star in two funnel streams. Closer to the star, the flow is
redirected by the octupolar component, with some of the matter
flowing towards the high-latitude poles, and the rest into the
octupolar belts. The shape and position of the spots differ from
those in a pure dipole case, where  crescent-shaped spots are
observed at the intermediate latitudes.

Simulations show that if the disk is truncated at the distance of
$r\approx 6.2 R_\star$ which is comparable with the co-rotation
radius, $r_{cor}\approx 6.8 R_\star$, then the high-latitude polar
spot dominates, but the accretion rate obtained from the
simulations (and from the accompanying theoretical calculations)
is about an order of magnitude lower than the observed one.  The
accretion rate matches the observed one if the disk is disrupted
much closer to the star, at $3.4 R_\star$. However, in that case
the octupolar belt spots strongly dominate. In the intermediate
case of $r\approx 4.3 R_\star$, the polar spots are sufficiently
bright, and the accretion rate is within the error bar of the
observed accretion rate, and this model can explain the
observations. However, an even better match has been obtained in
experiments with a dipole field twice as strong compared with one
suggested by D07.

The torque on the star from the disk-magnetosphere interaction is
small, and the time-scale of spin evolution, $2\times 10^7-10^9$
years is longer than the $2\times 10^6$ years age of V2129 Oph.
This means that V2129 Oph probably lost most of its angular
momentum in the early stages of its evolution, possibly, during
the stage when it was fully convective, and had a stronger
magnetic field.

The external magnetic flux
 of the star is strongly
influenced by the disk: the field lines connecting the disk and
the star inflate and form magnetic towers above and below the
disk. The potential (vacuum) approximation is still valid inside
the Alfv\'en (magnetospheric) surface
 where the magnetic stress dominates over the matter stress.

\end{abstract}


\section{Introduction}

Classical T~Tauri stars (CTTSs) are young, low-mass stars and
often show signs of a strong magnetic field (e.g.,
\citealt{basr92,john99,john07}). The magnetic field plays a
crucial role in disc accretion by disrupting the inner regions of
the disc and channeling the disk matter onto the star
(\citealt{ghosh78,ghosh79,koni91}). Zeeman measurements of the
surface magnetic fields of CTTSs show that the field is strongly
non-dipole in many stars
(\citealt{dona97,dona99,jard02,jard06,john07}).  The magnetic
field of the CTTS V2129 Oph has been recently measured with the
ESPaDOnS and NARVAL spectropolarimeters (D07). Decomposition of
the observed field into multipolar components has shown that the
field of V2129 is predominantly octupolar ($B_{oct}=1.2$ kG), with
a smaller dipole component ($B_{dip}=0.35$ kG) and a much smaller
contribution from the higher multipole components. The large-scale
magnetic field geometry has been calculated by extrapolating the
surface magnetic field to larger distances by suggesting that the
field is potential (vacuum approximation). The data presented in
D07 correspond to the June 2005 observations. More recent
measurements (July 2009) have shown that the field of V2129 Oph is
larger: $B_{oct}=2.1$ kG and $B_{dip}=0.9$ kG (\citealt{dona10},
hereafter D10). Those authors suggest that the field of V2129 Oph
varies in time due to the dynamo process. In this paper we assume
that the field is fixed and corresponds to that given by D07.
However, we do perform test experimental runs at larger values of
the dipole field.

Analysis of the chromospheric spot distribution in V2129 Oph have
shown that the polar, high-latitude spot dominates (D07, see also
\citealt{jard08}). D07 developed an approximate model where they
investigated possible paths of matter flow in the potential field
along closed field lines (see e.g., \citealt{greg06}). They
concluded that the disk should be disrupted
 by the dipole component far away from the star, at $r\approx 7 R_\star$.
In this case matter flows mainly along the field lines of the
dipole component and hits the star at high latitudes which
corresponds to observations. These estimates are useful
first-order approximations to the processes in V2129 Oph. As a
next step it is important to investigate matter accretion onto
V2129 Oph in global MHD
 simulations, where the assumption that the field is potential can be dropped,
and the interaction of the disk with the complex field of the star can be investigated.

In this paper, we developed a model star with parameters  of V2129
Oph and with magnetic field distribution close to that observed in
V2129 Oph (as described in D07). We approximated the field with
superposition of the main (dipole and octupole) moments, and
performed global 3D MHD simulations of matter flow around such a
star. In the past, we were able to model accretion onto stars with
quadrupole (\citealt{long07,long08}), and octupole \citep{long10a}
magnetic field components. In this paper we investigate for the
first time accretion onto a star with realistic magnetic field and
parameters corresponding to CTTS V2129 Oph. In the accompanying
paper, we investigate numerically accretion onto a model star with
parameters close to BP Tau \citep{long10b}.

The goals of this paper are the following:
(1)~ to investigate magnetospheric accretion onto this model star in global 3D MHD simulations,
(2)~ to investigate numerically the magnetic field structure around the star and
to compare it with the initial, potential field,
(3)~ to compare the simulated and observed accretion spots,
 (4)~to compare the results with results for
 a purely dipolar field,  (5)~to obtain the accretion rate from simulations and compare it
 with that obtained from observations, (6) to investigate
 the torque on the star.

Section~2 briefly describes the numerical model we use in our simulations. Section~3 compares the
observed and modeled magnetic fields of V2129 Oph.
Section~4 discusses the results of the simulations. Our conclusions
are summarized in Section~5.

\section{Numerical model and reference values}

We solve the 3D MHD equations with a Godunov-type code in a
reference frame rotating with the star using the ``cubed sphere"
grid. The model has been described earlier in a series of previous
papers (e.g., \citealt{kold02,roma04a,long10a}). Hence, we
describe it only briefly here.

\textit{Initial conditions.} A rotating magnetic star is
surrounded by an accretion disc and a corona. The disc is cold and
dense, while the corona is hot and rarefied, and at the reference
point (the inner edge of the disc in the disc plane),
$T_c=100T_d$, and $\rho_c=0.01\rho_d$, where the subscripts `d'
and `c' denote the disc and the corona. The disc and corona are
initially in rotational hydrodynamic equilibrium (see
\citealt{roma02} for details). An $\alpha$-type viscosity is
incorporated into the code and helps regulate the accretion rate.
The viscosity is nonzero only in the disc, that is, above a
certain threshold density (which is $\rho = 0.2$ in the
dimensionless units discussed below). We use $\alpha=0.04$ in this
work.

\textit{Boundary conditions.} At both the inner and outer
boundaries, most of the variables $A$ are set to have free
boundary conditions, ${\partial A}/{\partial r}=0$. On the star
(the inner boundary) the magnetic field is frozen into the surface
of the star, that is, the normal component of the field, $B_n$, is
fixed,  though all three magnetic field components may vary. At
the outer boundary, matter is not permitted to flow back into the
region.
The free boundary condition on the hydrodynamic variables at the
stellar surface means that accreting gas can cross the surface of
the star without creating a disturbance in the star or the flow.
This neglects the complex physics of the interaction of the
accreting gas with the star, which is expected to produce a shock
wave in the stellar atmosphere (e.g., \citealt{kold08}).

\textit{A ``cubed sphere" grid} is used in the simulations. The grid
consists of $N_r$ concentric spheres, where each sphere represents an inflated cube.
Each of the six sides of the
inflated cube has an $N\times N$ curvilinear Cartesian grid. The whole grid
consists of $6\times N_r\times N^2$ cells.
In our dipole+octupole model of V2129 Oph, the octupole component dominates and
a high radial grid resolution is needed near the star. We achieve this by choosing
the radial size of the grid cells to be 2.5 times smaller than the angular size in the region $r \lesssim 4 R_\star$,
while it is equal to the angular size in the outer region as in all our previous work.
The typical grid used in simulations
has $6\times N_r\times N^2 =
6\times150\times41^2$ grid cells.

\textit{Reference values.}
The simulations are performed in dimensionless variables $\widetilde{A}$. To
obtain the physical dimensional values $A$, the dimensionless values $\widetilde{A}$ should be multiplied by the
corresponding reference values $A_0$ as $A=\widetilde{A}A_0$.
These reference values include: mass $M_0=M_\star$, where $M_\star$ is the mass of the star;
distance $R_0=R_\star/0.35$,
where $R_\star$ is the radius of the star; velocity $v_0=(GM/R_0)^{1/2}$. The
 time scale is period of rotation at $r=R_0$: $P_0=2\pi R_0/v_0$ .

To derive the reference values for the magnetic field, we take
into account that the tilts of the dipole and octupole moments
relative to the rotational axis are small, and hence we use the
formulae for the axisymmetric case: $\mu_1=0.5
B_{1\star}R_\star^3$ and $\mu_3=0.25 B_{3\star} R_\star^5$ (see,
e.g., \citealt{long10a}). We suggest that the polar field:
$B_{1\star} = B_{dip}$, $B_{3\star} = B_{oct}$.
 The reference magnetic moments for the dipole and
octupole components are $\mu_{1,0}=B_0R_0^3$  and
 $\mu_{3,0}=B_0R_0^5$ respectively, where $B_0$ is the reference magnetic field.
Hence, the dimensionless magnetic moments are:
$\widetilde{\mu}_1=\mu_1/\mu_{1,0}$,
$\widetilde{\mu}_3=\mu_3/\mu_{3,0}$. We take one of the above
relationships; for example, the one for the dipole; to obtain

\begin{equation}
\label{eq-B0} B_0=\frac{\mu_{1,0}}{R_0^3} = \frac{0.5
B_{1\star}}{\widetilde{\mu_1}}\bigg(\frac{R_\star}{R_0}\bigg)^3=
7.5\bigg(\frac{B_{1\star}}{350 {\rm
G}}\bigg)\bigg(\frac{1}{\widetilde{\mu_1}}\bigg) {\rm G} .
\end{equation}
 Hence, the reference magnetic field depends on
the dimensionless parameter $\widetilde\mu_1$ and the dipole
magnetic field, $B_{1\star}$. The reference density
$\rho_0=B_0^2/v_0^2$ and the reference mass accretion rate
$\dot{M}_0=\rho_0 v_0 R_0^2$ also depend on these parameters:
\begin{equation}
\label{eq-rho0} \rho_0=B_0^2/v_0^2 = 1.5\times10^{-13}
\bigg(\frac{B_{1\star}}{350 {\rm
G}}\bigg)^2\bigg(\frac{1}{\widetilde{\mu_1}}\bigg)^2 \frac{\rm
g}{{\rm cm}^3} ,
\end{equation}
\begin{equation}
\label{eq-Mdot0} \dot{M}_0=\rho_0 v_0
R_0^2=1.0\times10^{-8}\bigg(\frac{B_{1\star}}{350 {\rm
G}}\bigg)^2\bigg(\frac{1}{\widetilde{\mu_1}}\bigg)^2
\frac{M_\odot}{{\rm yr}}.
\end{equation}

\noindent We find the other reference values in a similar way: the
torque $N_0=\rho_0v_0^2R_0^3$; energy flux,
$\dot{E}_0=\rho_0v_0^3R_0^2$; energy flux per unit area,
$\dot{F}_0=\rho_0v_0^3$; temperature $T_0=\mathcal{R}p_0/\rho_0$,
where $\mathcal{R}$ is the gas constant; and the effective
blackbody temperature $T_{\mathrm{eff,0}} = (\rho_0
v_0^3/\sigma)^{1/4}$, where $\sigma$ is the Stefan-Boltzmann
constant.  Tab. \ref{tab:refval-1} shows the reference values
which are fixed in all models. Tab. \ref{tab:refval-2} shows the
reference values for our main models which depend on the
parameters $\widetilde{\mu}_1$ and $B_{1\star}$ (see \S 4). In the
paper, we show our results in dimensional units. Formulae
\ref{eq-rho0}, \ref{eq-Mdot0} and Tab. \ref{tab:refval-2} help
scale the results to other parameters.

In most of the models we change the dimensionless octupolar moment
$\widetilde\mu_3$ in same proportion as $\widetilde\mu_1$ so as to
keep the ratio $\mu_3/\mu_1$ fixed (models 1-3). To find the ratio
between the dimensionless moments, we use the approximate formulae
for aligned moments: and obtain the ratio
$\widetilde{\mu}_3/\widetilde{\mu}_1=B_{3\star} \tilde R_\star^2/2
B_{1\star}\approx 0.22$ (where $\tilde R_\star=0.35$). In a few
test cases, we increased the relative strength of the dipole field
(models 4-5).

\begin{table}
\centering
\begin{tabular}{llllllll}
\hline
 Reference values:\\
 \hline
  {$M_\star(M_\odot)$}               & $1.35$                \\
  {$R_\star(R_\odot)$}               & $2.4$           \\
  {$R_0$ (cm)}                       & $4.8\e{11}$               \\
  {$v_0$ (cm s$^{-1}$)}              & $1.9\e7$                    \\
 {$P_0$ (days)}                     & $1.8$                       \\
\hline
\end{tabular}
\caption{The reference values which are fixed in all
models.}\label{tab:refval-1}
\end{table}

\begin{table*}
\centering
\begin{tabular}{llllllll}
\hline
Model & $\widetilde\mu_1$ &{$B_{1\star}$ (G)} & {$B_0$ (G)} & {$\rho_0$ (g cm$^{-3}$)}& {$\dot M_0$ ($M_\odot$yr$^{-1}$)}&{$F_0$ (erg  cm$^{-2}$s$^{-1}$)}&{$N_0$ (g cm$^2$s$^{-2}$)}\\
\hline
model 1 & 1.5               &$350$                  & $5$         &  $6.7\e{-14}$           &$4.7\e{-9}$  & $4.8\e{8}$  &   $2.8\e{36}$ \\
model 2 & 1                 &$350$                  & $7.5$       & $1.5\e{-13}$            & $1.0\e{-8}$   & $1.1\e{9}$  &  $6.2\e{36}$ \\
model 3 & 0.5               &$350$                  & $15$        &  $6.0\e{-13}$           &  $4.2\e{-8}$  & $4.4\e{9}$   & $2.5\e{37}$ \\
\hline
model 4 & 1.5               &$700$                  & $10$        &  $2.7\e{-13}$           &$1.9\e{-8}$  & $1.9\e{9}$  &   $1.1\e{37}$ \\
model 5 & 1.5               &$1050$                 & $15$        &  $6.0\e{-13}$           &$4.2\e{-8}$  & $4.4\e{9}$  &   $2.5\e{37}$ \\
\hline
\end{tabular}
\caption{The reference values which depend on the parameters
$B_{1\star}$ and $\widetilde\mu_1$.
 The dimensional
values can be obtained by multiplying the dimensionless values
obtained in different simulations by these reference values (see
sec. 2 for details). In models 1-3, we varied the parameter
$\widetilde\mu_1$, and in models 4 and 5, the parameter
$B_{1\star}$. }\label{tab:refval-2}
\end{table*}

\section{Reconstructed magnetic field of V2129 Oph and the field used in 3D simulations}

The magnetic field of V2129 Oph has been reconstructed from observations by D07.
The authors decomposed the measured surface magnetic field into its spherical harmonic components.
They then approximated the field as a superposition of low-order multipoles, adjusting the magnitudes
of the multipole moments and their orientations to give the best match of the modeled field
to the measured field.

In one of the best matches the surface magnetic field of V2129 Oph
is approximated by a 0.35~kG dipole and a 1.2~kG octupole field,
with corresponding magnetic moments tilted by $\Theta_1=30^\circ$
and $\Theta_3=20^\circ$ (towards phases 0.7 and 0.9)  with respect
to the rotational axis (D07).We choose this field distribution as
the one for our dipole+octupole model. Higher-order multipoles and
azimuthal component are also present, but they carry much less
magnetic energy than the two main components.

Note that method of field reconstruction from observations has
some uncertainties, and the error bars are different in different
stars and for different components. The error bar is typically
$\sim25\%$ for the strongest field component (the octupole in the
case of V2129 Oph) but larger for the weaker field components. In
V2129 Oph, the dipole field is 3-4 times weaker than the octupole,
and the error bar can be 3-4 times larger, that is
$\lesssim100\%$. The error bars on the phases and tilts of the
magnetic axes (about the rotational axis) also depend on the
actual values: the phases are less accurate when the tilts are
low. In the case of V2129 Oph, the error bar in the phase of the
dipole axis is likely to be 0.1-0.2 rotation cycle, or larger,
while the tilt of the dipole axis is accurate to no better than
$10^\circ-20^\circ$.

\begin{figure}
\begin{center}
\includegraphics[width=7.5cm]{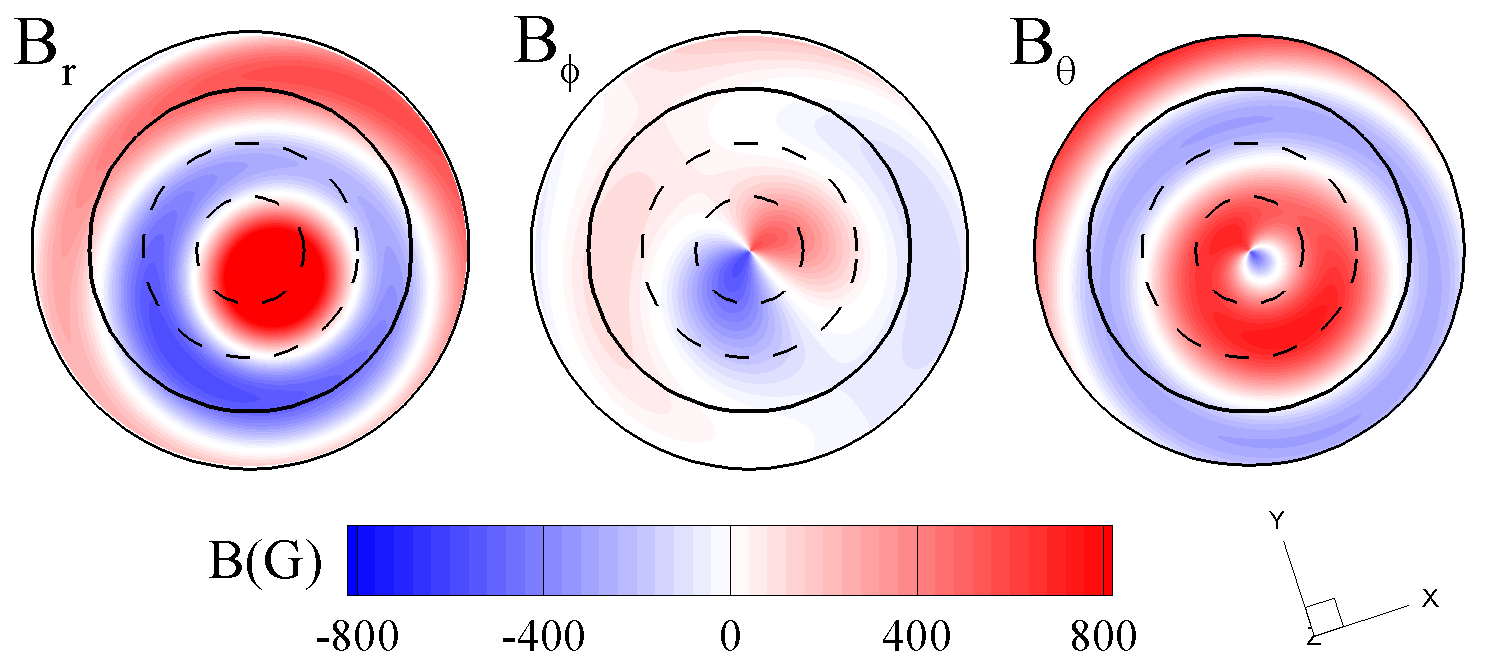}
\caption{\label{bcompv} Polar projections of the magnetic field
components at the stellar surface in the dipole+octupole model of
V2129 Oph in spherical coordinates: radial magnetic field, $B_r$;
azimuthal magnetic field, $B_\phi$; and meridional magnetic field,
$B_\theta$.  The outer boundary, the bold circle and the two inner
dashed circles represent the latitude of $-30^\circ$, the equator,
and the latitudes of $30^\circ$ and $60^\circ$, respectively. The
red and blue regions represent positive and negative polarities of
the magnetic field.}
\end{center}
\end{figure}

Fig. \ref{bcompv} shows the polar projection of the modeled (dipole plus octupole) field on the star's surface.
The style of the figures and the components of the field
$B_r$, $B_\phi$ and $B_\theta$  are similar to those reconstructed from observations by D07
(see their Figs. 12 and 13). Comparisons of our plots with the D07 plots show that the radial
(the strongest) component of the magnetic field is very similar in both plots (see top left panel in Fig. 12 of D07).
It shows a
strong positive pole, slightly misplaced relative to the rotational axis, then a negative octupolar ring
above the equator, and a positive octupolar ring below the equator. In the D07 plot,
the octupolar rings are not smooth in the azimuthal direction, which shows the presence of
the non-axisymmetric modes.
However, this difference is not very important for the present modeling, because most of the magnetic energy
is in the axisymmetric multipoles.

The  surface distribution of the meridional component of the
field, $B_\theta$ (Fig. \ref{bcompv}, right panel) is quite
similar to that in Fig. 13  of D07 (left bottom panel). In both
cases we see an inner ring of positive polarity (red), and an
outer ring of  negative polarity  (blue). The distribution of the
azimuthal component, $B_\phi$, shown in Fig. \ref{bcompv}, is
somewhat similar to that shown in Fig. 12 of D07 (left middle
panel). However, our azimuthal field is weaker because we do not
take into account the toroidal component of the field (see the
right panel of Fig. 13 of D07 for the toroidal component). We
suggest that the toroidal and the higher-multipole poloidal
components may influence the spot shapes near the star and can be
included in future modeling.

\begin{figure}
\begin{center}
\includegraphics[width=8.0cm]{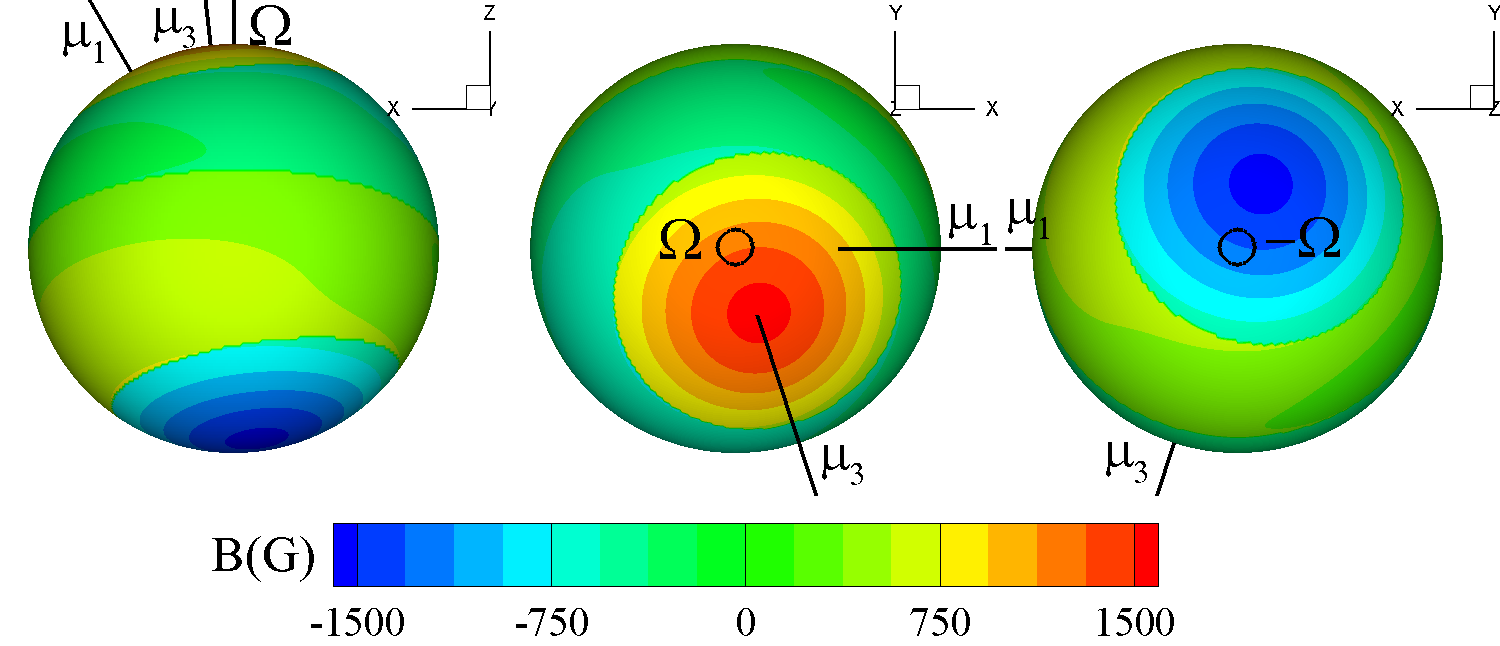}
\caption{\label{bsurfv} The surface magnetic field in the
dipole+octupole model of V2129 Oph ($\mu_1=1.5$, $\mu_3=0.33$,
$\Theta_1=30^\circ$, $\Theta_2=20^\circ$, $\phi_3=-72^\circ$) as
seen from the equatorial plane (left panel), the north pole
(middle panel) and the south pole (right panel). The colors
represent different polarities
 and strengths of the
magnetic field; red/yellow means positive, blue/dark green means negative.}
\end{center}
\end{figure}

\begin{figure*}
\begin{center}
\includegraphics[width=14.0cm]{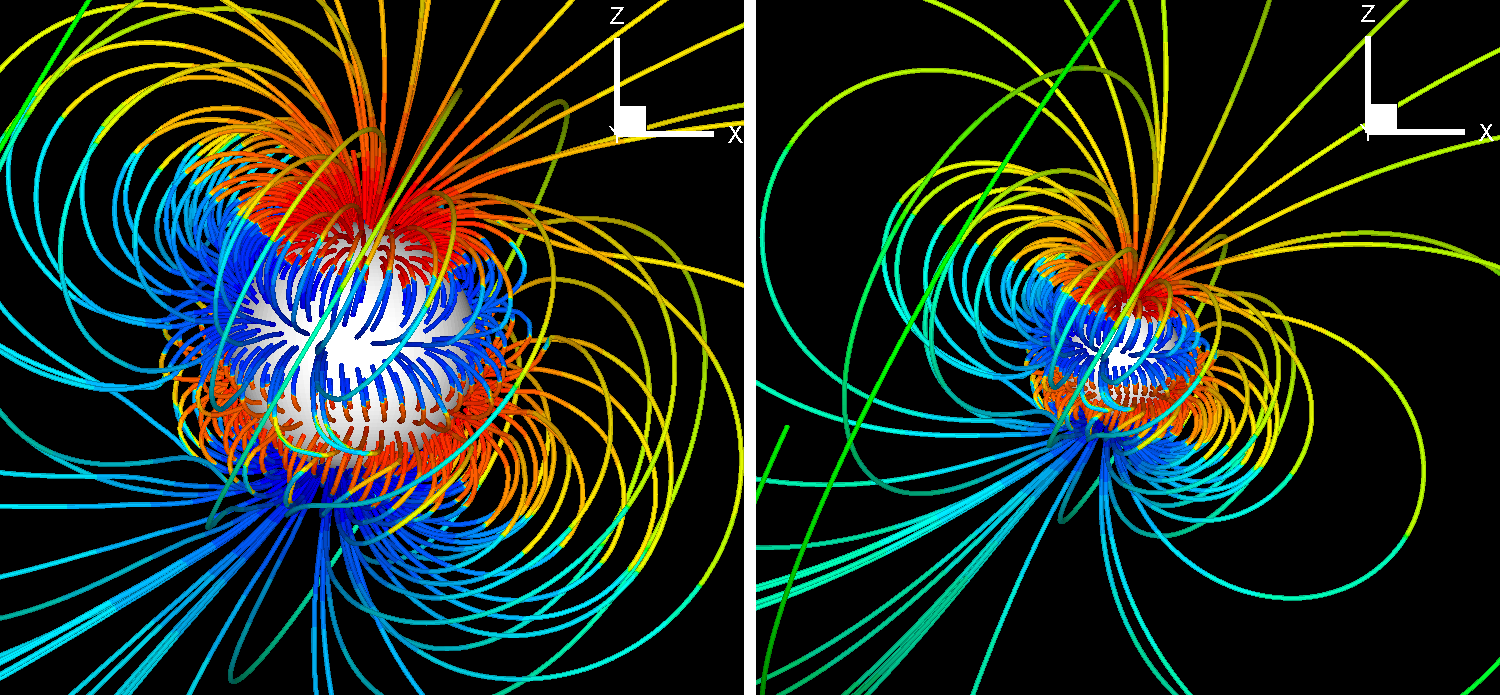}
\caption{\label{v2129-mag} The distribution of the magnetic field in the dipole+octupole model of V2129
at $t=0$.
The field is close to octupole near the star (left panel), but the dipole component dominates
at $r\gtrsim 2 R_\star$ (right panel).
The colors along the field lines represent
different polarities and strengths of the field.
}
\end{center}
\end{figure*}

Fig. \ref{bsurfv} shows the distribution of the modeled field on
the surface of the star. The middle and right panels show strong
northern (positive) and southern (negative) poles. The left panel
shows that there are also northern (negative) and southern
(positive) belts. One can see that the field is a typical octupole
field tilted by $20^\circ$ (see also \citealt{long10a}). The
dipole component is present but is much smaller.

An important radius is the radius where the dipole and octupole
components of the field are approximately equal. Below, we derive
an approximate value of such a radius, assuming that the dipole
and octupole moments are aligned with the rotational axis. The
magnetic fields of the dipole and octupole vary with distance as
(see \citealt{long10a}, Eqn. 1): (i) in the equatorial plane:~
$B_1={\mu_1}/{r^3}$, ~ $B_3=1.5 {\mu_3}/{r^5}$, and (ii) in the
polar direction:~ $B_1={2\mu_1}/{r^3}$, ~ $B_3={4\mu_3}/{r^5}$.
Equating $B_1$ and $B_3$ in the equatorial plane and evaluating
$\mu_1$ and $\mu_3$ at the surface of the star (e.g., $\mu_1=0.5
B_{1\star} R_\star^3$,  we obtain the distance where the dipole
and octupole components are equal:

\begin{equation}
\label{eq2} r_{eq}=R_\star
\bigg(\frac{3}{4}\frac{B_{3\star}}{B_{1\star}}\bigg)^{1/2} \approx
1.6 R_\star.
\end{equation}
  Here, we took into account that $B_{1\star}=0.35$ kG and $B_{3\star}=1.2$ kG.
That is, the octupolar field is expected to dominate at $r_{eq}\lesssim 1.6 R_\star$, whereas the dipole field
is expected to dominate at larger distances.

Fig. \ref{v2129-mag} shows the distribution of the intrinsic magnetic field in our dipole+octupole model
of V2129 Oph (at $t=0$). The left panel shows that the field near the star has
an octupolar structure up to a radius
of $r\sim (1.5- 2) R_\star$ which is in agreement with the $r_{eq}$ estimated above.
The right panel shows that at larger distances the field has a more ordered, dipolar structure.

\section{Modeling of accretion onto V2129 Oph}
\label{sec:results}

In simulating accretion onto CTTS V2129 Oph, we use its  mass and
radius derived from observations which are $M_\star=1.35 M_\odot$,
$R_\star=2.4 R_\odot$ (see D07 and references therein) and
rotation period $P_\star=6.53$ days (see \citealt{shev98}). These
values imply that the corotation radius $r_{cor}=6.8 R_\star$. An
estimate age of V2129 Oph is $2\times10^6$ years (D07).

A number of simulation runs were performed. To change  the size of
the magnetosphere (which is equivalent to the changing of the
dimensional accretion rate at the fixed magnetic field on the
star), we varied the parameter $\widetilde{\mu}_1$ from 0.5 to 5
and obtained a set of simulation runs where the disk stops at
different distances from the star. We find from the simulations
that at $\widetilde{\mu}_1\gtrsim 3$ the inner disk radius is
larger than the corotation radius and the disk moves away due to
the ``propeller" effect. At  smaller parameters,
$1.5\lesssim\widetilde{\mu}_1\lesssim 3$ the magnetosphere is
large, $r_t > (6-8) R_\star$, however, the accretion rate is much
smaller than the observed one. We chose some intermediate case
with  $\widetilde\mu_1=1.5$ as the main one. In that case the disk
stops far away from the star $r_t\approx 6 R_\star$ (which is
close to the corotation radius, $r_{cor}\approx 6.8 R_\star$), and
at the same time an accretion rate is not too low. We refer to
this case as the main one. It corresponds to model 1 in Tab.
\ref{tab:mdot}. For this case
 we performed detailed analysis of different features of the flow, hot spots and the large-scale
 magnetic field evolution (see sections 4.1-4.3, 4.5, 4.6).
 In models 2 and 3 (see Tab. \ref{tab:mdot}) we investigate
 additional cases, where the disk comes closer to the star and accretion rate is higher
 and is closer to the observed one. However, the low-latitude octupolar belt spots become
 stronger than the polar spot
 (see sec. 4.7).
In models 4 and 5 we keep the same octupolar field, but increase
the dipole magnetic field by a factor of 2 and 3 respectively.
 This  helps to increase an accretion rate and at the same time
octupolar belt spots are weak and the spot shape is closer to the
observed one (see sec. 4.9).
 In model 6 we investigate accretion
 on to a star with a pure dipole field, and compare results with the main case (see sec. 4.4).
 Models 7-9 are
 theoretical models (see sec. 4.8).

\subsection{Matter flow around the star}
\label{sec:matflow}

\begin{figure}
\begin{center}
\includegraphics[width=7.5cm]{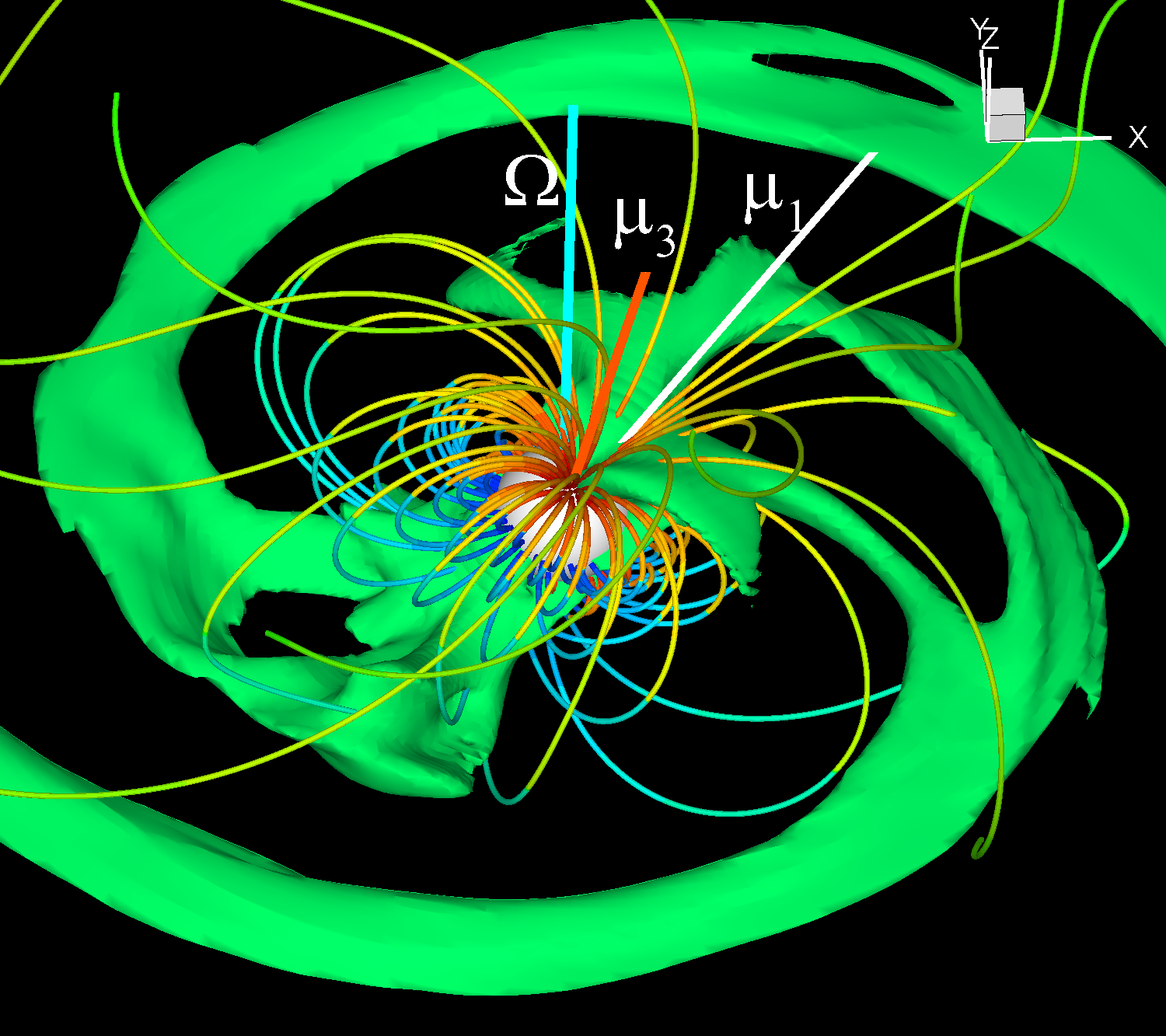}
\caption{\label{3dv} 3D view of matter flow onto the star in the
dipole+octupole model of V2129 Oph (model 1). The disc is shown by
a constant density surface in green with $\rho=6\times 10^{-15}$ g
cm$^{-3}$. The colors along the field lines represent different
polarities and strengths of the field. The thick cyan, white and
orange lines represent the rotational axis and the dipole and
octupole moments respectively.}
\end{center}
\end{figure}

Here we discuss results of simulations in our main case (model 1
in Tab. \ref{tab:mdot}). Initially, we place the inner edge of the
disk at the radius $r = 8 R_\star$. The small viscosity
incorporated into the disk leads to inward matter flow. First, the
disk moves inwards over $t\approx 4$ rotations (Keplerian
rotations at $r=1$). Later, the funnel streams form and the matter
accretes onto the star quasi-stationary. We were able to reach
time  $t=8-10$ in this type of simulations. Fig. \ref{3dv} shows
the matter flow and selected field lines at $t=9$. One can see
that the disk is stopped by the dipole component of the field, and
matter flows towards the star from two directions, which
correspond to the directions of the dipole component tilt
($\Omega\mu_1$ plane).

\begin{figure}
\begin{center}
\includegraphics[width=8.0cm]{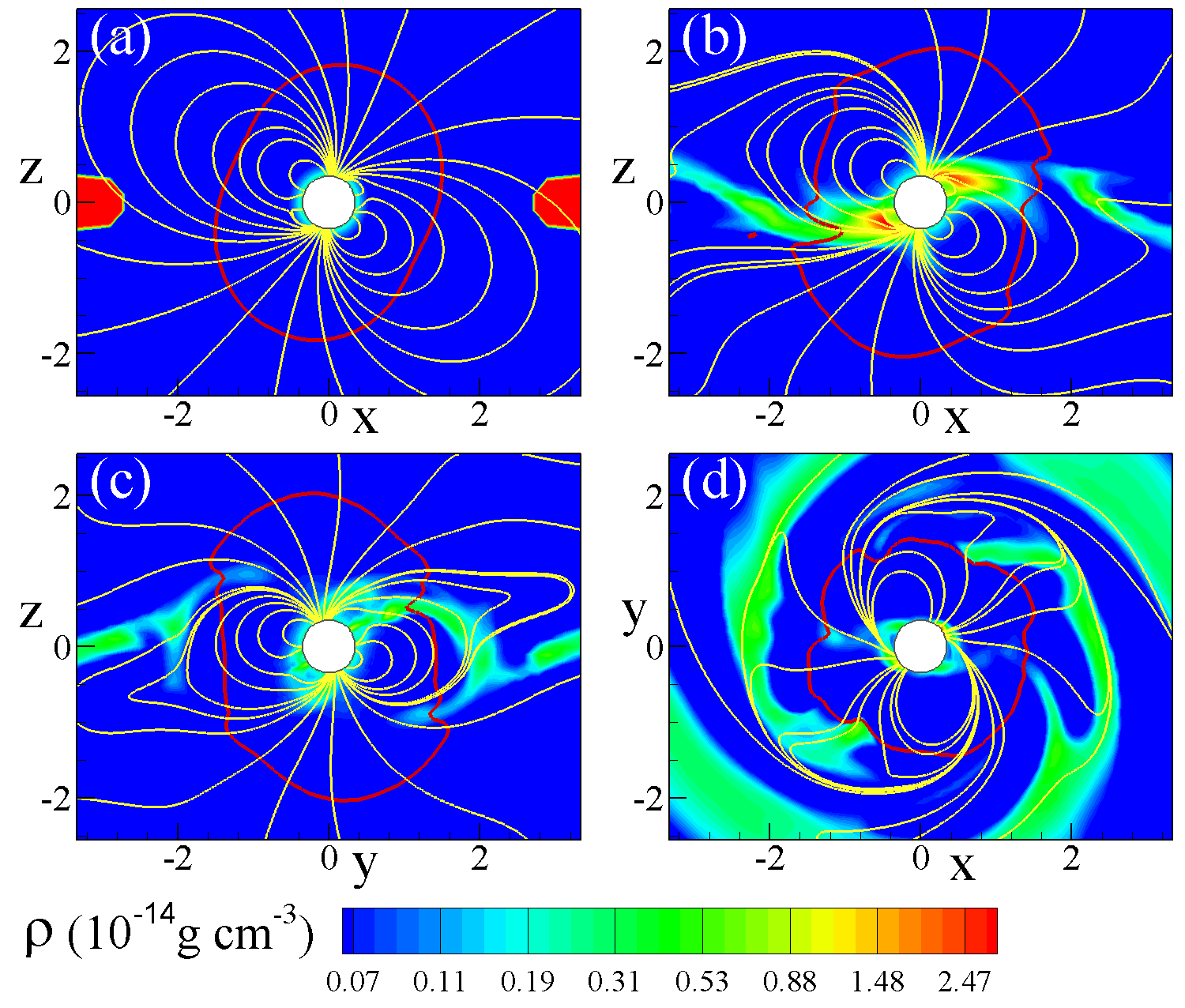}
\caption{\label{flowv} Density distribution in different slices at
$t=9$ for model 1. Panels (a) and (b) show $xz$ slices at $t=0$
and $t=9$. Panels (c) and (d) show $yz$ and $xy$ slices at $t=9$.
The red line shows the distance at which the matter stress equals
the magnetic stress.}
\end{center}
\end{figure}

Fig. \ref{flowv} shows the density distribution in three slices,
$xz$, $yz$ and $xy$. One can see that the disk is  disrupted by
the dipole component of the field far away from the star, at $5
R_\star \lesssim r_t \lesssim 7 R_\star$, and the funnel streams
form at $r\approx 6 R_\star$. The red line shows the surface where
the matter and magnetic field stresses balance each other. The
dominant stress is given by the $\phi\phi$-component of the stress
tensor. The magnetospheric radius $r_m$ is then the radius where
$\beta\equiv(p+\rho v^2)/(B^2/8\pi)=1$. Inside this radius the
magnetic stress dominates (\citealt{ghosh78,ghosh79,long10a}; see
also \citealt{bess08} for different definitions of the inner disk
radius). We obtain $r_m\approx (4-5) R_\star$.

Closer to the star, the octupole component of the field becomes stronger,
and each funnel stream splits into two parts, one of which flows towards the
magnetic pole, and the other into an octupolar belt. This produces two types of accretion
spots on the star. We discuss this in more detail in \S \ref{sec:spots}.

\begin{figure*}
\begin{center}
\includegraphics[width=14.5cm]{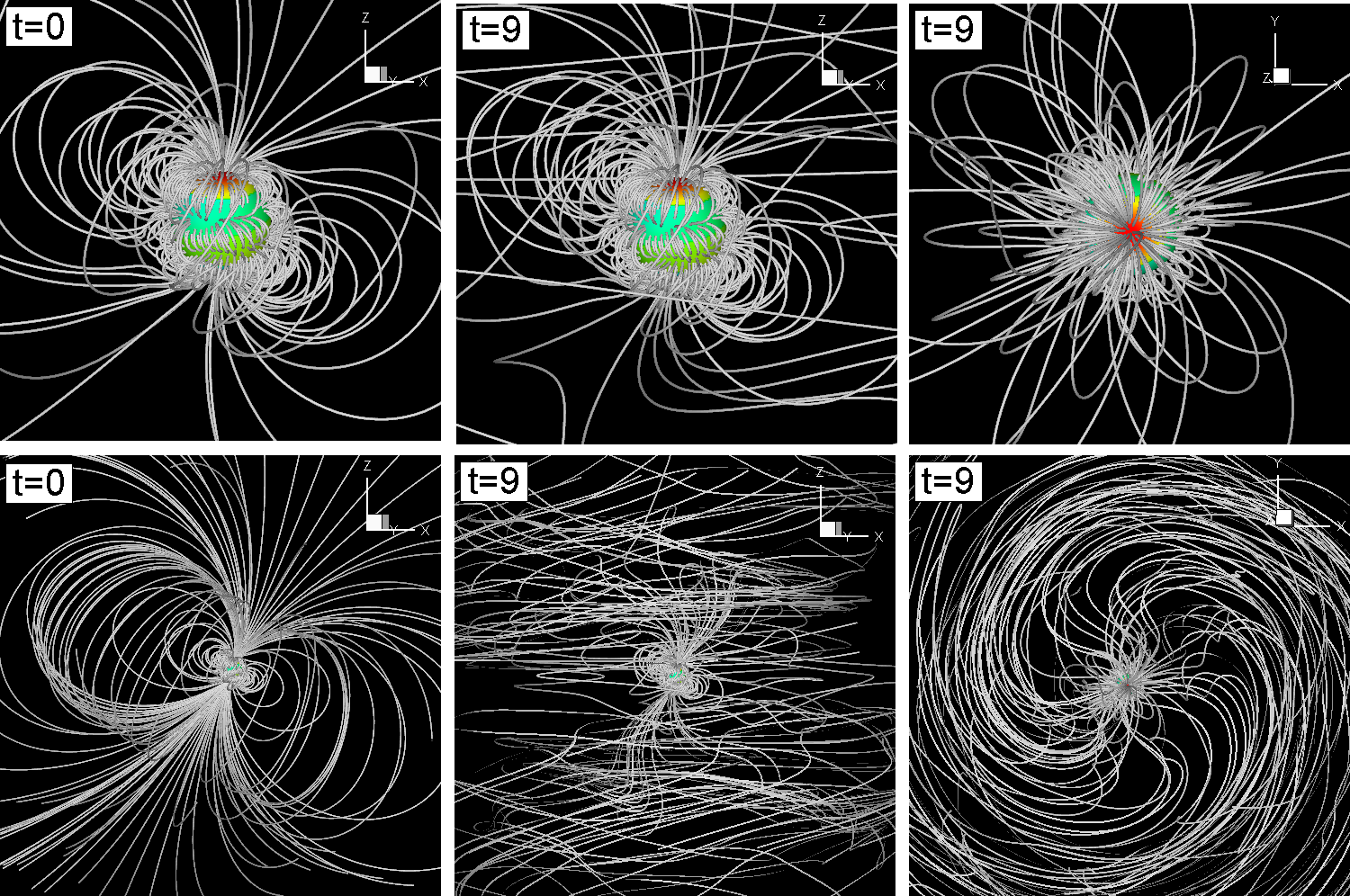}
\caption{\label{blinesv} Comparison of the initial (potential) magnetic field distribution
at $t=0$ (left panels) with the field distribution at $t=9$ (middle and right panels).
The top panels show the magnetic field in the vicinity of the star, while the bottom panels
show the field distribution in the whole simulation region.
The left and middle columns show the side view, while the right column shows the axial view of the field.}
\end{center}
\end{figure*}

\subsection{Magnetic field structure and the validity of the potential approximation}
\label{sec:magfield}

Next, we investigate the  magnetic field evolution due to the
disk-magnetosphere interaction. Fig. \ref{blinesv} (left two
panels) shows the initial magnetic field distribution near the
star (top panel) and in a larger region (bottom panel). These
plots correspond to the {\it potential approximation} which is
often used for extrapolating the surface magnetic field to larger
distances (e.g., D07). As the simulations proceed, the plasma and
magnetic field motions  create currents and fields in the
simulation region and the resulting magnetic field geometry
strongly departs from the potential one. The dominant processes
are connected with the differential rotation of the foot-points of
the magnetospheric field lines which have different angular
velocities at the star's surface, in the disk, and in the corona.
Differential rotation leads to conversion of the rotational energy
into  magnetic energy and to inflation of the field lines (e.g.,
\citealt{aly80,love95,roma98,agap00}). The foot-points threading
the star usually rotate faster than those threading the disc, and
the magnetic field lines wrap around the rotational axis forming a
magnetic tower (e.g. \citealt{lynd96}). Formation of such a tower
has been observed in different numerical simulations (e.g.,
\citealt{usty00,kato04,roma04b,roma09,usty06}) and is a natural
outcome of disk-magnetosphere interaction. The middle and right
bottom panels of Fig. \ref{blinesv} show formation of such a
tower. So, at large scales the magnetic field strongly departs
from the potential one.

The situation is different near the star.  Fig. \ref{blinesv} (top
panels) shows that the field structure at $t=9$ is almost the same
as at $t=0$. This is the region $r \lesssim (4-5) R_\star$, where
the magnetosphere is not disturbed by the disk, and magnetic
stress dominates over the matter stress, and hence the magnetic
fields produced by plasma flow are insignificant compared with the
main magnetic field of the star produced by currents flowing
inside the star. The potential approximation is valid in this
region.

We conclude that the potential approximation used in many works to recover the external
 magnetic field from the
observed one (e.g. \citealt{jard02}; D07) is sufficiently good in
the vicinity of the star, where the magnetic stress dominates over
the matter stress (inside the Alfv\'en surface, $r< r_m$), and
where the magnetosphere is not disturbed by the accretion disc.
However, the field strongly departs from the potential one at
larger distances.

It is often suggested that in young stars a strong stellar wind
stretches the stellar field lines in the radial direction, like in
the solar corona, and this determines the geometry of the external
magnetic field  (e.g. \citealt{safi98,greg06,jard08}). We should
note that this solar-type scenario can dominate in the case of
diskless, weak-line T Tauri stars. However, if a large-scale
magnetic field truncates the accretion disk, then a different,
magnetic tower geometry dominates. Such a tower can accelerate a
small amount of coronal matter up to high velocities and can carry
some angular momentum from the star to the corona. However, both
processes are significant only in cases of rapidly rotating stars
(e.g., \citealt{roma05,usty06}), and are less important in slowly
rotating stars, like V2129 Oph \citep{roma09}.

\citet{greg08} noted that the dominance of the octupolar component
near the star causes a smaller fraction of the stellar surface to
be connected to open field lines compared with the case of a
purely dipolar field. We observed that in V2129 Oph the amount of
flux in open field lines depends on  the position of the disk and
on the strength of the dipole component which dominates at
$r\gtrsim 2R_\star$ and which determines the amount of open flux.
The field lines inside the inner disk are closed. All other field
lines of the dipole component
 inflate and represent open magnetic flux. The octupolar component contributes towards open flux in polar region,
but this flux is much smaller than the dipole open flux, as
suggested by \citet{greg08}.

\subsection{Accretion spots in V2129 Oph}
\label{sec:spots}

\begin{figure}
\begin{center}
\includegraphics[width=8.0cm]{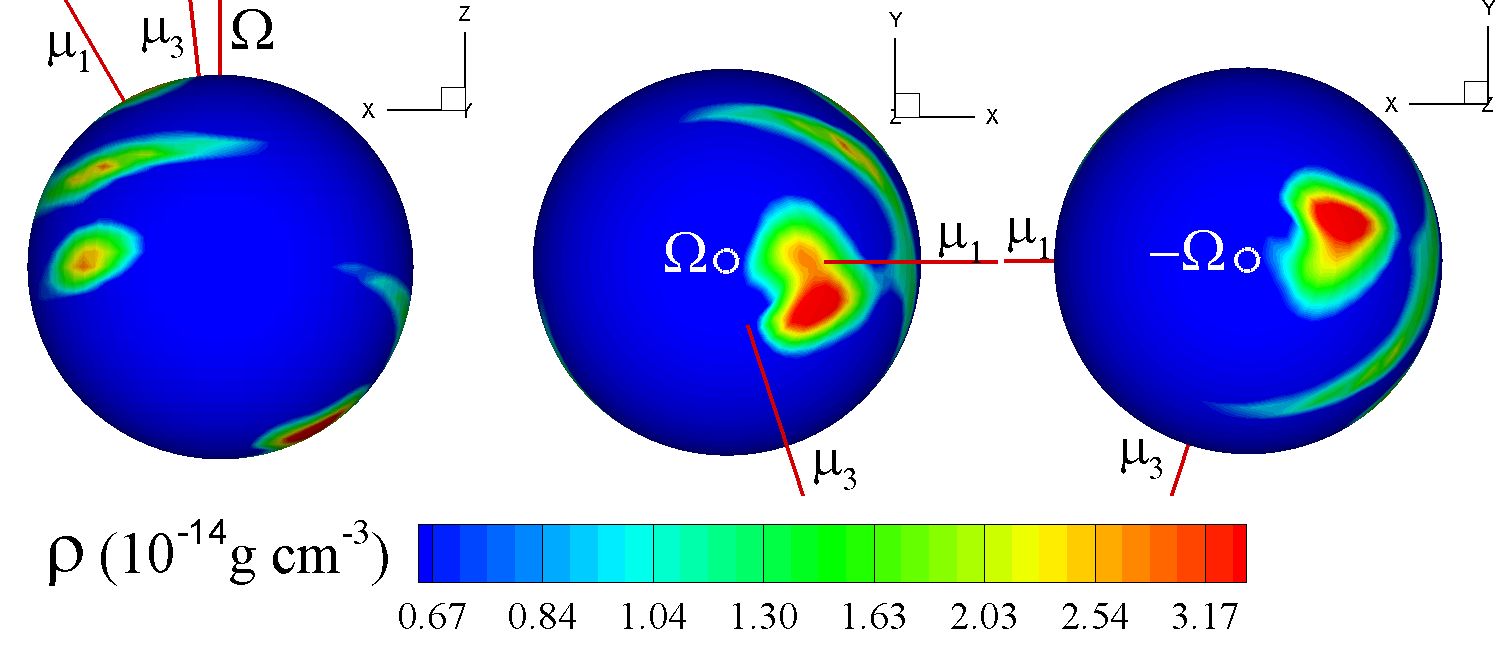}
\caption{\label{hsv} The calculated hot spots viewed from different directions
at $t=9$: from the equatorial plane (left-hand panel); from the
north pole (middle panel); and from the south pole (right-hand panel). The color contours show
the density distribution of the matter. Solid lines
represent the magnetic moments of the dipole ($\mu_1$) and octupole ($\mu_3$).}
\end{center}
\end{figure}

\begin{figure}
\begin{center}
\includegraphics[width=8.0cm]{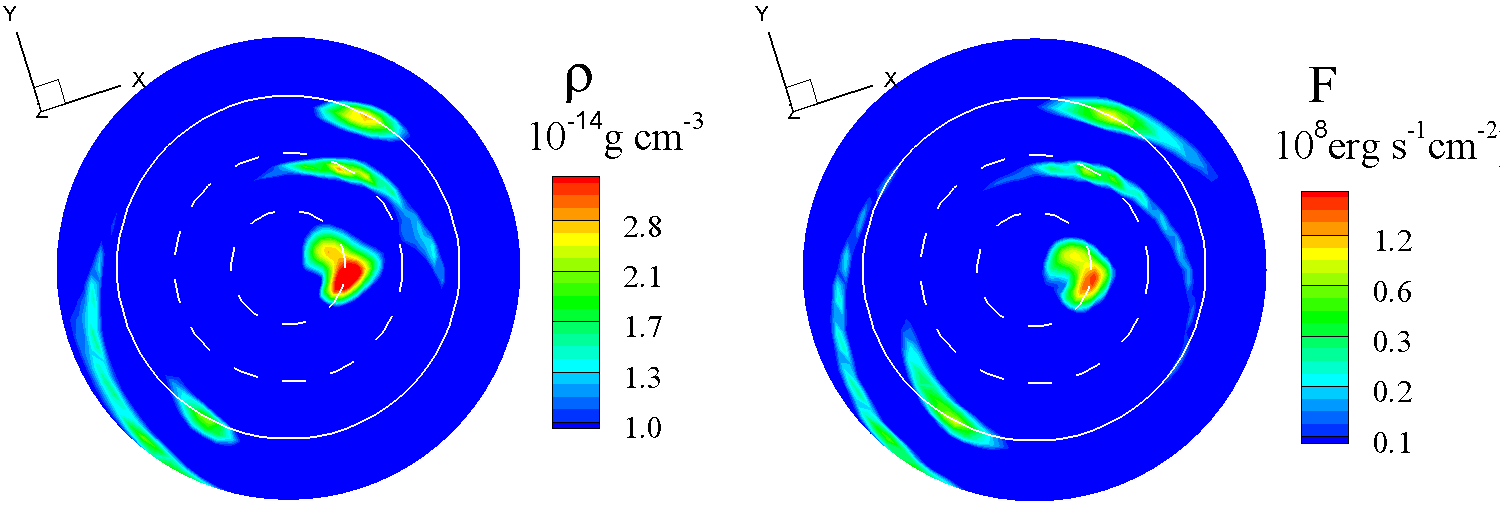}
\caption{\label{hspv} The calculated hot spots in a polar projection down to $120^\circ$
from the north pole. Left
panel: density distribution; right panel: energy flux distribution.
The equator is shown as a solid line.
The dashed lines
represent the latitudes $30^\circ$ and $60^\circ$ respectively.}
\end{center}
\end{figure}

\begin{figure}
\begin{center}
\includegraphics[width=6.0cm]{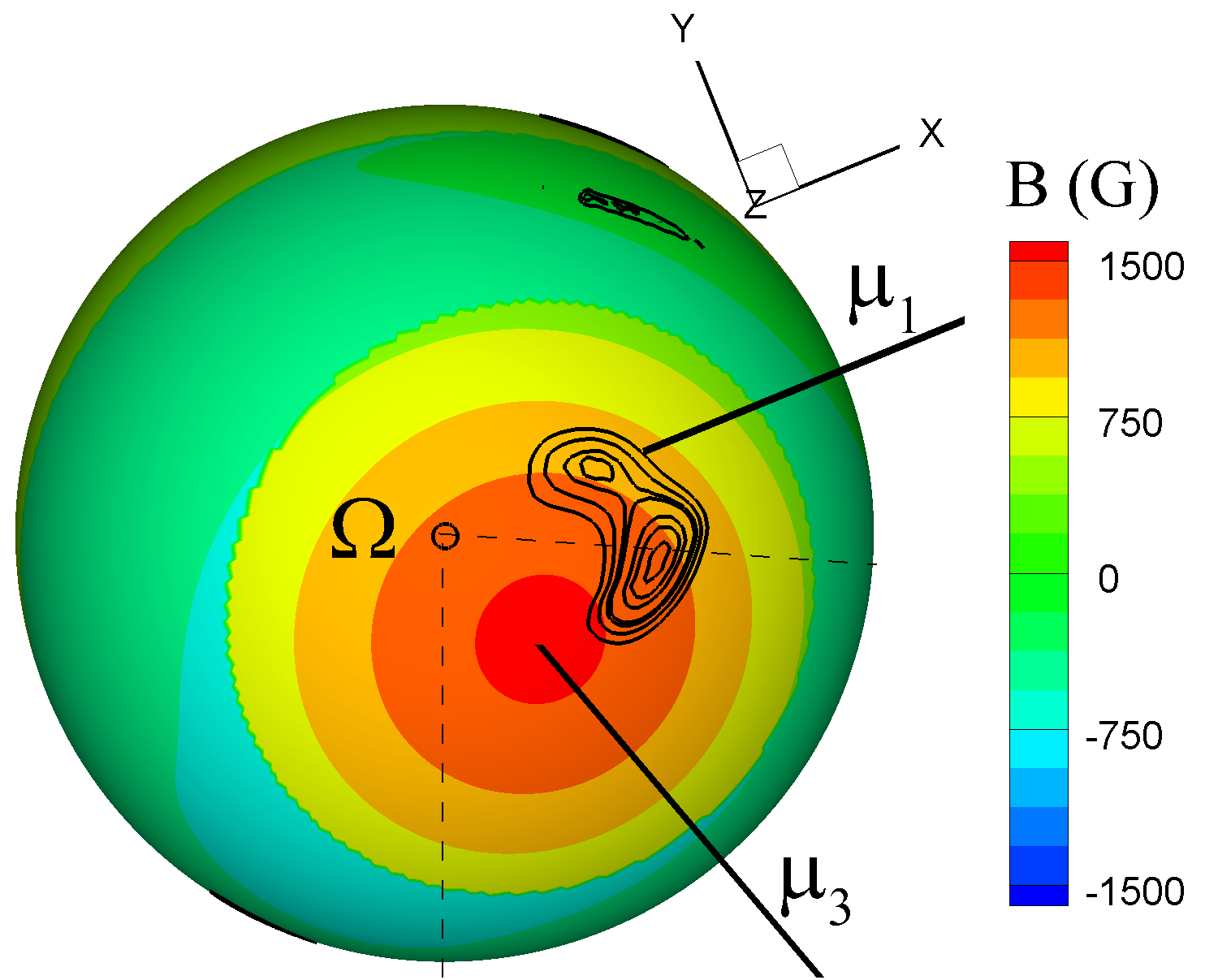}
\caption{\label{spot-b} The color background shows the distribution of the total \emph{}magnetic field
on the surface of the star. The black contours show the energy distribution in the accretion flow
at the star. The black solid lines show the direction of the dipole and octupole moments. The dashed line
shows the direction towards the brightest part of the spot. The vertical dashed line shows
the origin line for the phase count in D07 paper (clockwise).}
\end{center}
\end{figure}

D07 analyzed  the brightness enhancements observed in V2129 Oph in
the chromospheric lines, like in CaII IRT, and also in the HeI D3
line, which presumably forms in the shock wave near the star's
surface. The right panel of Fig. 9 of D07 shows  the CaII IRT
excess emission relative to the basal chromospheric emission,
which can be interpreted as the chromospheric ``hot spot". Fig. 9
of D07 shows that the main part of the spot, the polar spot, is
located at colatitudes $0^\circ<\theta_c<50^\circ$ with a center
at $\theta_c\approx 25^\circ$. The position of this spot almost
coincides with the position of the magnetic pole (see Fig. 12 top
left panel of D07). In addition, non-axisymmetric spots of lower
brightness are seen at higher colatitudes up to $\theta_c\approx
90^\circ$.

In our numerical model, matter flows towards the star in funnel streams. We suggest that all matter
which approaches the star freely moves inward through the star, and hence we neglect the complex
physics of the stream-star interaction.
Instead, we calculate the matter and energy fluxes
carried towards the star, and our accretion spot represents a slice in density/energy distribution
across the funnel stream.

In reality, the stream
interacts with the dense regions of the chromosphere
and photosphere, and forms a shock wave where gas is heated, and the energy is radiated
 mainly behind the shock front (e.g., \citealt{lamz95,calvet98}). The shock wave may
  oscillate due to cooling/heating instability
behind the shock front \citep{kold08}, and the stream-star
interaction may lead to formation of waves on the star's surface
and possible outflows (\citealt{cran08,cran09}; see also
\citealt{bric10}). We suggest that the heating processes in the
dense layers of star's atmosphere may excite such lines as CaII
IRT, HeI D3 and others, and hence we expect a correlation between
the position of the chromospheric spots observed by D07, and
accretion spots observed in our simulations.

Fig. \ref{hsv} shows different views of the density spots. One can see that there are two
main spots, the polar ones, which almost coincide with the octupolar magnetic poles.
There are also two elongated spots
at much lower latitudes where matter flows towards the magnetic octupolar belts.

Fig. \ref{hspv} shows a polar view of the spots. The left panel
shows the density distribution. The right panel shows the
distribution of the energy flux which is calculated as:
$F=\rho\bm{v}\cdot\hat{r}[(v^2-v_\star^2)/2+{\gamma
p}/(\gamma-1)\rho]$ (here $v_\star$ is the surface velocity of the
star)
 and shows the location of the
main energy deposition in the funnel stream. One can see that the
polar density spot is located at colatitudes of
$5^\circ\lesssim\theta_c\lesssim 45^\circ$ and is centered at
$\theta_c\approx 30^\circ$. The octupolar belt spots are
azimuthally elongated and are located at
$60^\circ\lesssim\theta_c\lesssim 80^\circ$. The energy spots have
similar shapes and locations. We also calculated the distribution
of the matter flux $\rho v_p$ and temperature $T$ (calculated in
the suggestion that all kinetic energy is converted into black
body radiation). We observed, that their distribution is similar
to that of the density and energy fluxes shown in Fig. \ref{hspv}.
Any of these values can be used for analysis of accretion spots on
the star. Note that the size of spots in our simulations depends
on the chosen density/energy level, so that the densest/brightest
parts of spots have a smaller size.

Comparison of the accretion spots obtained in our simulations with the CaII IRT chromospheric ``spots"
observed in V2129 Oph (Fig. 9 of D07)
shows that the polar spot is present in both cases and
has a similar location and shape.
The simulated spot is somewhat more compact than the observed one
(see however \S 3.6 where we discuss the dependence of the
spot size on the density/energy levels). Also, the center of the observed spot is located
at a slightly higher latitude than the simulated spot. In both cases, the polar spots
are located near the magnetic poles seen in Fig. 1 of our paper (and Fig. 12 of D07),
though the position of the simulated density spot is somewhat different in phase
compared with the magnetic pole position.

Fig. \ref{spot-b} shows that in our simulations an accretion spot
(shown in black contour lines) does not coincide with the maximum
of the magnetic field (red region on the star). A spot is leading
(about 0.15 in phase). Here we note that the spot is located
between the dipole and octupolar poles. We suggest that the main
accretion stream moves toward the dipole magnetic pole and hence
the accretion spot is located close to this pole. However, an
octupole field also influences the accretion stream, and as a
result a spot is located between the dipole and octupolar magnetic
poles. This may explain why we observe the phase difference of
about 0.15 between the main (octupolar) magnetic pole and
accretion spot. The uncertainty in the tilt and phase of the
dipole field derived from observations is about 10\%. If we assume
that the phase difference between the dipole and octupolar poles
is smaller, then we expect that the accretion spot will be closer
to the octupolar magnetic pole (as observed by D07). There is also
another possibility, which is that the funnel stream is dragged by
the rapidly rotating disk and moves slightly ahead of the more
slowly rotating magnetosphere, and this leads to the phase
difference between the spot position and the magnetic pole (as
shown in simulations by \citealt{roma03,roma04a}).

\begin{figure*}
\begin{center}
\includegraphics[width=11.0cm]{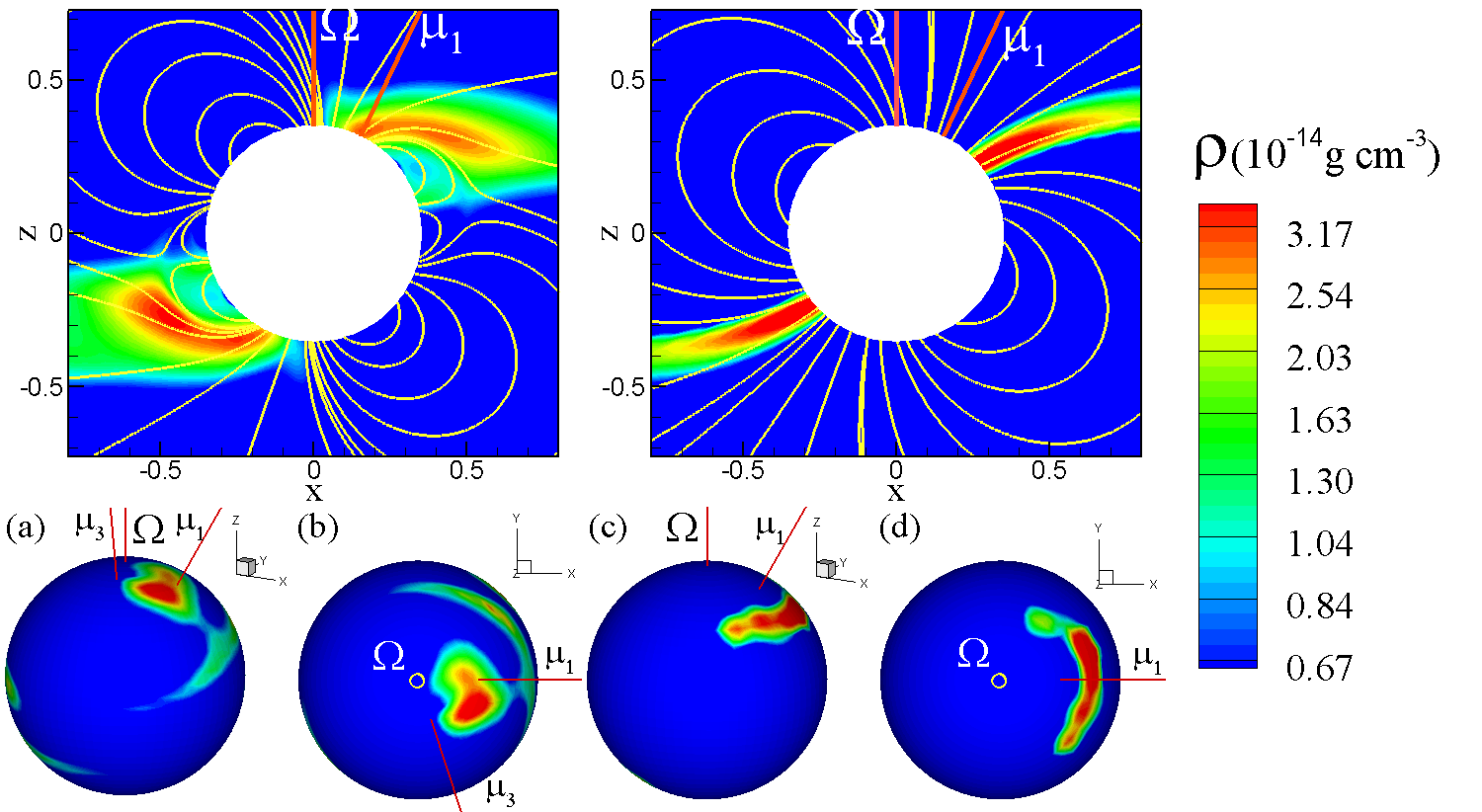}
\caption{\label{v2129_comp} Comparison of matter flow and hot
spots in  the dipole+octupole model of V2129 Oph (left panels)
with a pure dipole case (right panels) at $t=9$. The top panels
show the density distribution in the $xz-$plane (color background)
and sample field lines. The bottom panels show the density
distribution in the hot spots as seen from along the $y-$axis
(panels a,c) and along the rotational axis (panels b,d).}
\end{center}
\end{figure*}

\subsection{Comparison with a pure dipole case (model 6)}
\label{sec:dipole}

We performed a special set of runs where the magnetic field of
V2129 Oph is approximated with the dipole component only (zero
octupolar field). The strength of the dipole field,  0.35 kG, and
its tilt, $\Theta=30^\circ$, are the same as in the main
dipole+octupole case. Figure \ref{v2129_comp} (top right) shows
the matter flow and spots in the $xz$-plane in the case of a pure
dipole field. One can see that matter flows in two narrow streams
and the hot spots have the azimuthally elongated, crescent shapes,
typical for stars with the dipole field \citep{roma04a,kulk05}.
The spots are centered at a much lower latitude, $\theta_s\approx
45^\circ$, compared to spots in the dipole+octupole case.

The left panels show the $xz$ cross-section and spots in our main dipole+octupole case for comparison.
One can see that initially matter
flows along the dipole field lines. However, closer to the star, the octupole component
determines the flow pattern: the funnel stream  becomes wider and splits into a polar stream and
an equatorial belt stream. Note that the polar stream is redirected by the octupolar component
to much higher latitudes compared with the stream in a pure dipole case.

We conclude that the high-latitude round accretion spots obtained in simulations of the
dipole+octupole model are in much better qualitative agreement with the
observed accretion spots on V2129 Oph (D07) than the lower-latitude
crescent-shaped accretion spots obtained in the pure-dipole model.


In this work we have neglected the higher-order multipoles which
can dominate near the stellar surface. We should note that they
cannot change the position of the main accretion spot, but instead
can only redistribute the density inside the main spot.

\subsection{Area covered by hot spots}
\label{sec:areaspots}

\begin{figure}
\begin{center}
\includegraphics[width=7.0cm]{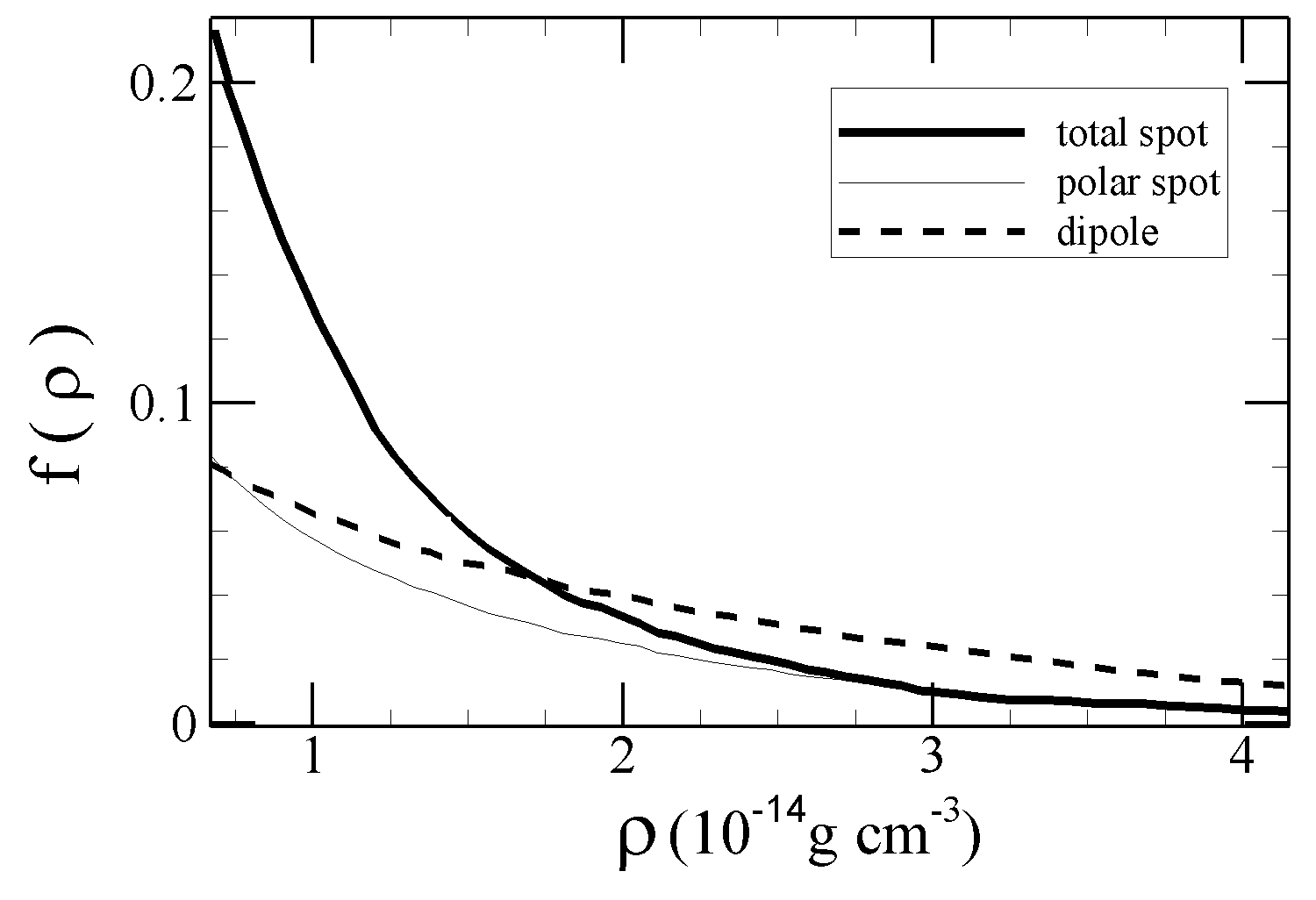}
\caption{\label{spotsarea} Fraction of the star's surface covered by spots of density $\rho$ and higher at $t=9$.
The bold line shows the spot coverage in the dipole+octupole model of V2129 Oph where both polar and octupolar belt spots are
present.  The thin solid line shows same, but for the polar spot only. The dashed
line shows same but for a pure dipole model of V2129 Oph.}
\end{center}
\end{figure}

Here we calculate the area covered by the accretion spots. We
choose a density $\rho$, and calculate the area $A(\rho)$ covered
by the part of the spots with density higher than $\rho$, and the
fraction of the star covered by these spots: $f(\rho)=A(\rho)/4\pi
R_\star^2$. We calculate this fraction for different values of
$\rho$. To separate the contribution from the polar spots, we
calculate separately the fraction of the star covered by all spots
($f_{tot}$), and by polar spots only, $f_{polar}$.  We also
calculate the fraction of the star covered by spots in the pure
dipole case, $f_{dipole}$. Fig. \ref{spotsarea} shows how the spot
coverage varies with $\rho$. One can see that the area covered by
spots in the dipole case is larger than that in the
dipole+octupole case if  $\rho>1.7\times 10^{-14}$g cm$^{-3}$
($\widetilde{\rho}>0.26$). The situation is reversed if the
density level is smaller.
 Note that in earlier comparisons of accretion to stars with dipole and quadrupole fields,
we observed a similar trend: at higher density levels, the spots
in the pure dipole case are larger than those in the dipole
 plus quadrupole cases, while at lower density levels, the situation is reversed \citep{long10a}.

\begin{figure*}
\begin{center}
\includegraphics[width=14.0cm]{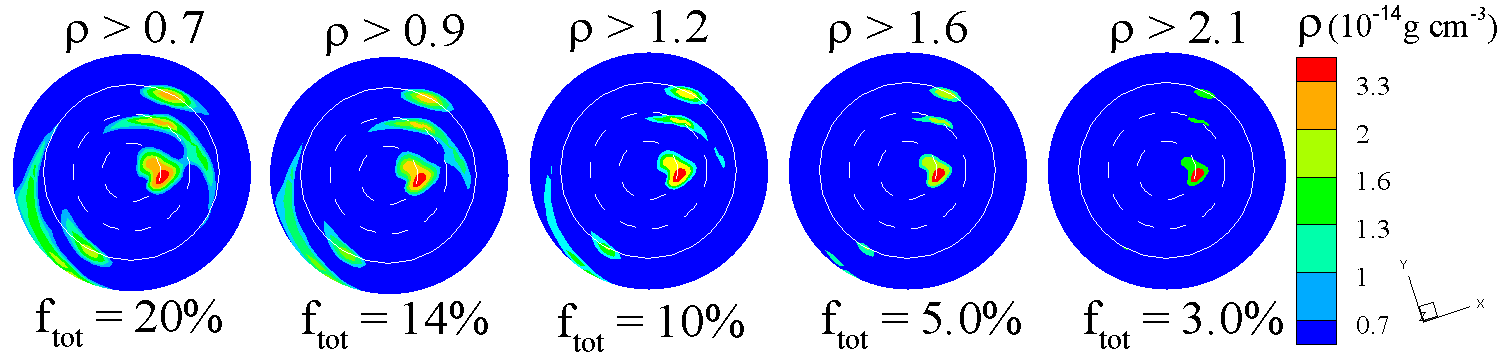}
\caption{\label{4spots} Example of spots with a cut at the
different density levels $\rho$ and corresponding spot coverage
for the full spot $f_{tot}$.}
\end{center}
\end{figure*}

In another figure, Fig. \ref{4spots}, we choose several density
levels and show the size of spots corresponding to these density
levels. One can see that at the lowest density level,
$\rho=7\times 10^{-15}$g cm$^{-3}$ ($\widetilde{\rho}=0.1$), the
spots occupy $f_{tot}\approx 20\%$, while at the highest density
level, $\rho=2.1\times 10^{-14}$g cm$^{-3}$
($\widetilde{\rho}=0.3$) , they occupy $f_{tot}\approx 3\%$. If
only the polar spot is taken, then the coverage for the same
density levels is smaller
--- $f_{polar}\approx 8.0\%$ and $f_{polar}\approx 2.3\%$
respectively. Fig. \ref{spotsarea} shows that at even higher
density levels, the spot coverage is very small. For example, at
$\rho=3\times 10^{-14}$g cm$^{-3}$, $f_{tot}\approx 1\%$.

Thus, the spot size strongly depends on the density level. This
density distribution shows the structure of the funnel streams:
the density is higher in the middle of the stream, and decreases
toward the periphery. The matter flux, kinetic energy flux and
temperature have a similar spot-centered distribution and hence
the spot coverage is different at different energy/temperature
levels with the hottest spots occupying a smaller fraction of the
star's surface (see also \citealt{roma04a}).

The size of the accretion spots reconstructed from observations,
e.g., of the CaII line, also depends on brightness as seen in
Figs. 9 and 11 of D07 and Fig. 6 of D10. The fraction of the star
covered with each of visible spots is approximately 5\% which is
about 10\% for both spots (observations take into account only the
spots in the visible hemisphere, and do not take into account a
probable antipodal spots; to compare simulations with
observations, we multiply the observed fraction by a factor of
two). One can see that the brightest parts of the observed spots
cover a much smaller area of the star. The position and shapes of
the observed and simulated spots are similar: both are round and
are located at the co-latitude of $\theta_c\approx 30^\circ$.

Fig. \ref{4spots} (middle panel) shows that at this spot coverage
($f_{tot}=10\%$) the polar part of the spot strongly dominates,
while the octupolar belt spots are weaker. We conclude that there
is a good correspondence between the sizes and longitudinal
positions of the main, brightest parts of the spots obtained in
observations and simulations. However, there is a difference in
shapes and phases of the lower-density/energy parts of the spots
(see also \S4.4 and \S4.6).

It was predicted that the area covered by spots in the case of
complex magnetic fields should be smaller compared with the spot
area in the pure dipole case (e.g., \citealt{moha08,greg08}, see
also \citealt{calvet98}). In addition, \citet{calvet98} derived
from their model and observations of a number of CTTSs that their
spot coverage is expected to be very small, less than a percent in
most stars. However, our simulations show that the spot coverage
is higher than that predicted by \citet{calvet98}. It is possible
that these authors analyzed only the central,
 brightest parts of the spots. This problem requires further investigation.

\begin{table*}
\centering
\begin{tabular}{l@{\extracolsep{0.2em}}l@{}llllllllll}
\hline &  Observations of V2129 Oph ~~& $\widetilde\mu_1$ &
$\widetilde\mu_3$ & $B_{1\star}$(G) & $B_{3\star}$(G) &
$\widetilde{\dot M}$
& $\dot{M}$(M$_\odot$yr$^{-1}$)  & $r_t/R_\star$ &  spots (at $f=10\%$)& \\
\hline
~ &        &$-$  &$-$    & 350  & 1200    & $-$     & $1.2\e{-9}$ (max)   & $-$    & polar& \\
~ &       &$-$  &$-$    & 350  & 1200    & $-$     & $6.3\e{-10}$  & $-$           & polar& \\
~ &        &$-$  &$-$    & 350  & 1200    & $-$     & $3.2\e{-10}$ (min)   & $-$   & polar& \\
\hline
   &  Model of V2129 Oph ~~& &  & &&
&   & &  \\
\hline
model 1.~ & dip+oct 3D &  1.5~~ &0.33   & 350  & 1200   & 0.015  & $7.2\times 10^{-11}$   &  6.2 & polar+small arcs & \\
model 2.~ & dip+oct 3D &  1.0~~ &0.22  & 350  & 1200    & 0.029  & $3.0\times 10^{-10}$  & 4.3   & polar+long arcs& \\
model 3.~ & dip+oct 3D &  0.5~~ &0.11  & 350  & 1200    & 0.023  &$1.0\times 10^{-9} $    & 3.4  & polar+rings& \\
\hline
model 4.~ & dip+oct 3D &  1.5~~ & 0.165   & 700  & 1200   & 0.029  & $5.2\times 10^{-10}$  & 5.9  & polar &\\
model 5.~ & dip+oct 3D &  1.5~~ & 0.11    & 1050 & 1200   & 0.036    & $1.5\times 10^{-9}$ & 5.5 & polar&\\
\hline
model 6.~ & dip 3D     &  1.5~~& 0.00  & 350  & 0    & 0.046   & $2.2\times 10^{-10}$   & 6.0 & crescent &  \\
\hline
model 7.~ & dip (theory) &$-$  &$-$   & 350  & 0    & $-$  & $6.2\times 10^{-11}$  ~~ & 6.2  & -&\\
model 8.~ & dip (theory) &$-$  &$-$   & 350  & 0    & $-$  & $2.2\times 10^{-10}$  ~~ & 4.3  & -&\\
model 9.~ & dip (theory) &$-$  &$-$   & 350  & 0    & $-$  & $5.1\times 10^{-10}$  ~~ & 3.4  & -&\\
\hline

\end{tabular}
\caption{The Table combines different observational properties of
V2129 Oph (top three rows), and shows different models used in
simulations and their results (see text). The dimensionless
parameters $\widetilde\mu_1$  and $\widetilde\mu_3$ determine the
dimensionless properties of the magnetosphere; $B_{1\star}$(G) and
$B_{3\star}$(G) are the polar dipole and octupole fields given by
observations; $\widetilde{\dot M}$ is the dimensionless parameter
of the accretion rate derived from simulations;
 $\dot{M}$(M$_\odot$yr$^{-1}$) is the dimensional accretion rate;  $r_t/R_\star$ is the truncation
 radius of the disk given in units of stellar radius. The last column describes the properties
 of spots at such a density/energy level when they cover about $f=10\%$ of the stellar surface.
The top three rows show the observed accretion rates including its
maximum, minimum and middle values. Model 1 is the ``main"
(reference) model and it is described in detail in different
subsections of sec. 4. It is also used for comparisons with other
models. Models 2 and 3 show results of simulations in cases where
accretion rate is higher, and the disk comes closer to the  star
(see sec. \ref{sec:accrate}). Models 4  and 5 show results of
simulations in cases, where the dipole component is twice and
thrice as
 large compared with observations (see sec. \ref{sec:mod4and5}).
Model 6 shows result of simulations in a pure dipole case (see
sec. \ref{sec:dipole}). Models 7-9 describe results of the
theoretical model (see sec. \ref{sec:accrate}).}\label{tab:mdot}
\end{table*}

\subsection{Rotational modulation}
\label{sec:rotmodulat}

\begin{figure}
\begin{center}
\includegraphics[width=7.0cm]{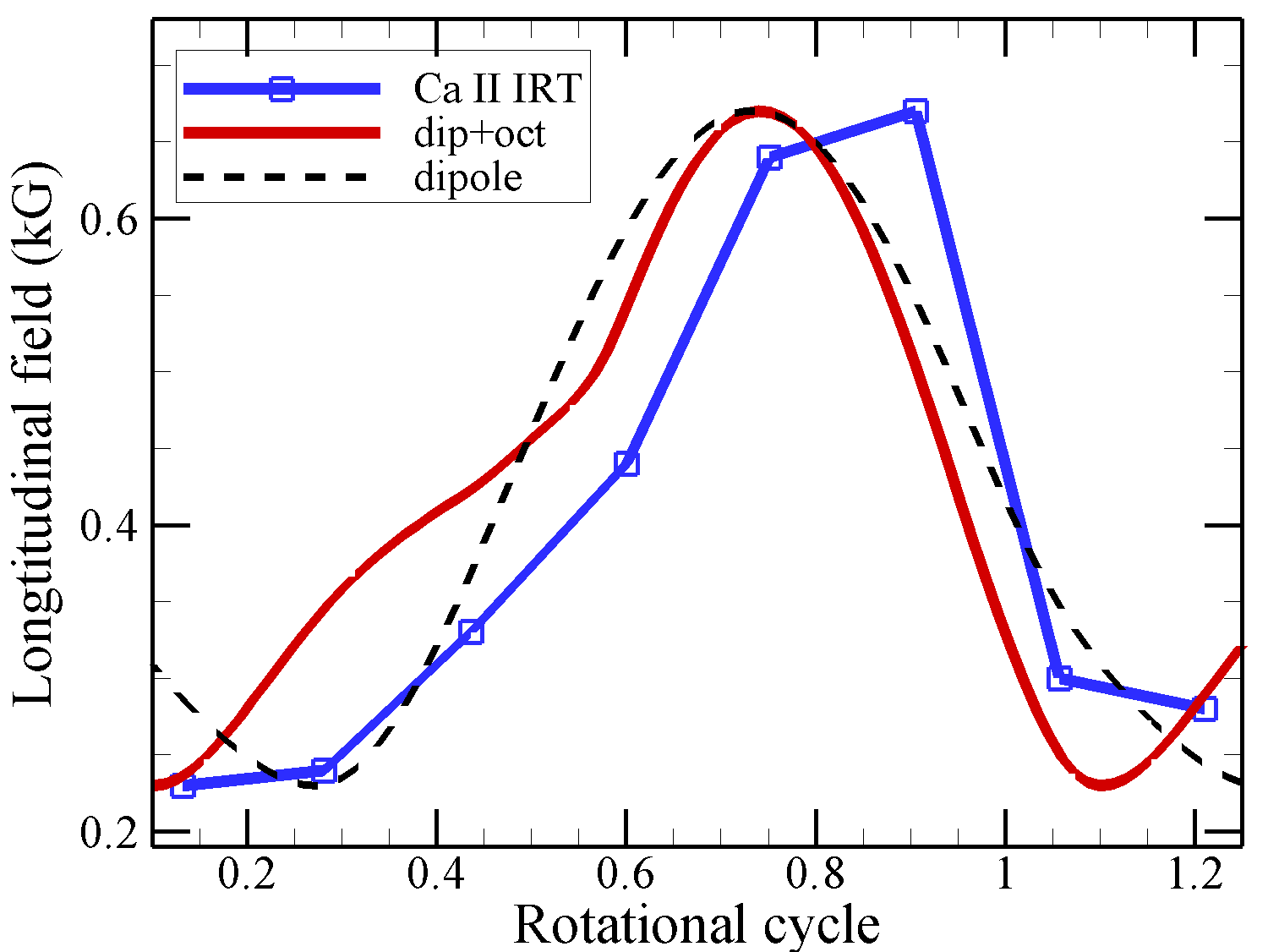}
\caption{\label{lcv} The rotationally modulated longitudinal
magnetic field derived by D07 from observations of the emission
line CaII IRT (blue line) compared with the rotationally modulated
radiation from spots obtained in our simulations (red line) and
from simulations of the pure dipole field (dashed black line) for
inclination angle $i=45^\circ$. }
\end{center}
\end{figure}

\begin{figure}
\begin{center}
\includegraphics[width=8.0cm]{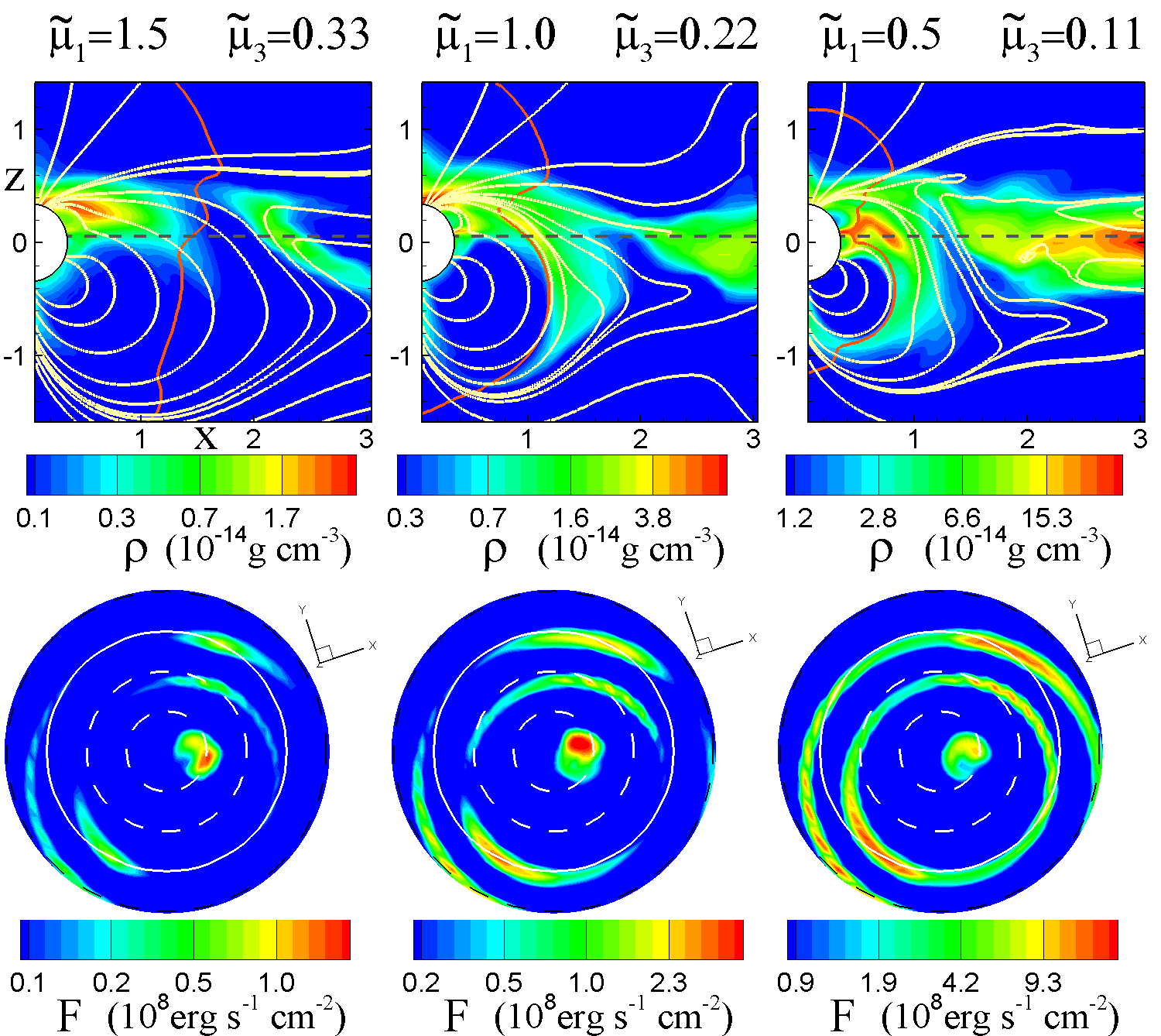}
\caption{\label{dip_10} Result of simulations at different magnetospheric sizes (parameters
$\widetilde\mu_1$) for models 1, 2, 3 (see Tab. \ref{tab:mdot}). Top panels show density distribution (background), selected field lines
(yellow lines), and $\beta=1$
line (red line) in the xz-slices.
Bottom panels show the energy flux distribution on the  star in the polar projection.
}
\end{center}
\end{figure}

D07 observed that the fluxes and amplitudes of the Zeeman signatures in the
  CaII IRT and HeI D3 emission lines exhibited strong
rotational modulation over the rotational cycle of the star which
is close to the 6.53 d period in V2129 Oph. The time-dependent
longitudinal field derived from these emission lines also shows
the rotational modulation (see Fig. 5 of D07). The emission in
CaII IRT traces the regions of higher excitation of this line in
the chromosphere, while the HeI D3 line is thought to form in the
shock wave \citep{beri98}. These features can be compared with the
rotationally modulated accretion spots obtained in our
simulations.

We calculate the rotationally modulated radiation from accretion
spots towards the observer located at an angle $i$ relative to the
rotational axis of the star. To calculate the radiation, we
suggest that all the energy flux of the funnel stream is converted
into isotropic black-body radiation in the shock wave (see details
in \citealt{roma04a}). We choose some late moment of time ($t=9$)
when the position of the spots does not vary much and obtain the
rotationally modulated radiation from such spots.

Fig. \ref{lcv} compares the rotationally modulated longitudinal
magnetic field derived from  CaII IRT observations by D07 (blue
line) with our rotationally modulated spot curves (red lines)
which we calculated at the inclination angle $i=45^\circ$ (D07).
 One can see the gradual rise and sharper drop in both simulated and observed curves which may be
a sign of the spots asymmetry. We calculated a similar
rotationally modulated curve for V2129 Oph with the magnetic field
approximated by a pure dipole field and found that the curve is of
a sinusoidal shape (see the dashed black line in Fig. \ref{lcv}).
We suggest that the non-sinusoidal shape may be connected with the
presence of the low-latitude (octupolar belt) part of the spot in
our simulations, and with the low-latitude components of the spots
in the observations.

The observed and simulated curves have a phase difference of about
0.1-0.2 which probably reflects the phase difference between the
simulated and observed spots.

\subsection{Accretion rate onto V2129 Oph}
\label{sec:accrate}

The accretion rate in V2129 Oph has been estimated by D07 using
the CaII IRT line to be $\dot M\approx 4 \times 10^{-9}$ M$_\odot$
yr$^{-1}$ (using the approach of  \citealt{moha05}). Using fluxes
from several emission lines and more recent empirical relations
between line emission fluxes and accretion luminosities
\citep{fang09}, a revised mass accretion rate was derived from the
same spectra: $\log (\dot M /(M_\odot \mbox{ yr}^{-1})) = -9.2\pm
0.3$, or: $\dot M = 6.3\times 10^{-10} M_\odot \mbox{ yr}^{-1}$
with the maximum and minimum values of $\dot M_{\rm max}  =
1.2\times 10^{-9} M_\odot \mbox{yr}^{-1}$ and $\dot M_{\rm
min}=3.2\times 10^{-10}M_\odot \mbox{yr}^{-1} $.

Below we compare the observed accretion rate with the rates
obtained from simulations and derived from theoretical
estimations. The dimensional accretion rate is
\begin{eqnarray}
\label{eq3} \dot M_{\rm simul} &=& \widetilde{\dot M} {\dot M_0} =
1.1\times10^{-8}~\frac{\rm{M_{\odot}}}{\rm
yr}~\bigg(\frac{\widetilde{\dot M}}
{{\widetilde\mu_1}^2}\bigg) \bigg(\frac{B_{* eq}}{175 {\rm G}}\bigg)^2 \nonumber\\
       &&\bigg(\frac{M_*}{1.35 M_{\odot}}\bigg)^{-\frac{1}{2}}
\bigg(\frac{R_*}{2.4 R_{\odot}}\bigg)^{\frac{5}{2}} ~.
\end{eqnarray}
In the case of $\widetilde\mu_1=1.5$,  we take $\widetilde{\dot
M}\approx 0.015$ from the simulations, and $\dot M_0$ from Tab.
\ref{tab:refval-2} and obtain a dimensional accretion rate  $\dot
M_{\rm simul}=7.2\times 10^{-11} M_\odot \mbox{yr}^{-1}$. This
accretion rate is about 9 times smaller than the observed
accretion rate (see Tab. \ref{tab:mdot}).

In an attempt to  resolve this issue, we performed simulation runs
at smaller values of the magnetospheric parameter,
$\widetilde{\mu}_1 = 1$ and $0.5$ (see models 2 and 3 in Tab.
\ref{tab:mdot}). In each case, we adjusted the dimensionless
octupolar moment using the ratio:
$\widetilde\mu_3/\widetilde\mu_1=0.22$, thus fixing the ratio
between the dipole and octupole moments. We observed from the
simulations that for $\widetilde{\mu_1}=1$, the accretion rate is
higher and it matches the minimum value derived from observations,
$\widetilde{\dot M}_{min}$ (see model 2 in the Table). For
$\widetilde{\mu_1}=0.5$, the accretion rate is sufficiently high
to match the observed accretion rate (see model 5).

Fig. \ref{dip_10} (top panels) shows  matter flow at different
$\widetilde{\mu}_1$. One can see that the size of closed
magnetosphere systematically decreases with $\widetilde{\mu_1}$.
We determine (very approximately) the truncation radius as the
radius at which the last closed (or, strongly disturbed)  field
line crosses the equatorial plane (shown as a dashed line on the
plot). Then for $\widetilde{\mu}_1=1.5, 1, 0.5$, we obtain a set
of radii,  $r_t\approx 6.2, 4.3, 3.4$. Each of these radii has an
error of about $\pm(0.1-0.3)$, because there are usually a few
field lines which can be considered as slightly disturbed lines.
Nevertheless, we see a clear trend of $r_t$ decreasing with
$\widetilde{\mu}_1$.

Fig. \ref{dip_10} (bottom panels) shows that an area covered with
the octupolar belt spots increases when $\widetilde{\mu}_1$
decreases, because the disk is truncated closer to the star, where
the octupolar component dominates. Note that in the case of
$\widetilde{\mu}_1=1$, the polar spot dominates over octupolar
ring spots if we take the brightest/densest $10\%$ of the stellar
surface. However, in the case of $\widetilde{\mu}_1=0.5$,
octupolar ring spots give the brightest part at any density/energy
level. Such octupolar spots have not been observed by D07. So, the
intermediate case ($\widetilde{\mu}_1=1$, model 2) can possibly
explain the observations, though the truncation radius $r_t\approx
4.3$ is somewhat smaller than the corotation radius, $R_{cor}=6.8
R_\star$.

To further investigate this issue, we  estimated the accretion
rate using a theoretical approach. Prior simulations done in our
group show that the formula for the Alfv\'en radius
\citep{elsn77}:
\begin{equation}
\label{eq4} r_t=k(GM_\star)^{-1/7}\dot{M}^{-2/7}\mu_1^{4/7}.
\end{equation}
is approximately applicable  for the truncation of the accretion
disks, if $k\approx 0.5$ (\citealt{long05}; see also
 \citealt{bess08} for different approaches for finding the inner disk radius).
We can rewrite  this formula in the convenient form:

\begin{eqnarray}
\label{eq5} \dot M_{\rm theor} &=& 6.9\times 10^{-11}
\left(\frac{k}{0.5}\right)^\frac{7}{2}\left(\frac{r_t}{6
R_{\star}}\right)^{-\frac{7}{2}}
\left(\frac{B_{1\star}}{350 {\rm G}}\right)^2 \cdot \nonumber\\
&&\left(\frac{M_*}{1.35
M_{\odot}}\right)^{-\frac{1}{2}}\left(\frac{R_*}{2.4
R_{\odot}}\right)^{\frac{5}{2}}
 ~ \frac{\rm{M_{\odot}}}{\rm yr}~.
\end{eqnarray}


\noindent Tab. \ref{tab:mdot} (models 7-9) shows theoretically
derived values of $\dot M$ for the truncation radii $r_t$ obtained
in models 1-3.  One can see that these accretion rates
approximately correspond to those obtained in the 3D simulations.

\begin{figure}
\begin{center}
\includegraphics[width=8.0cm]{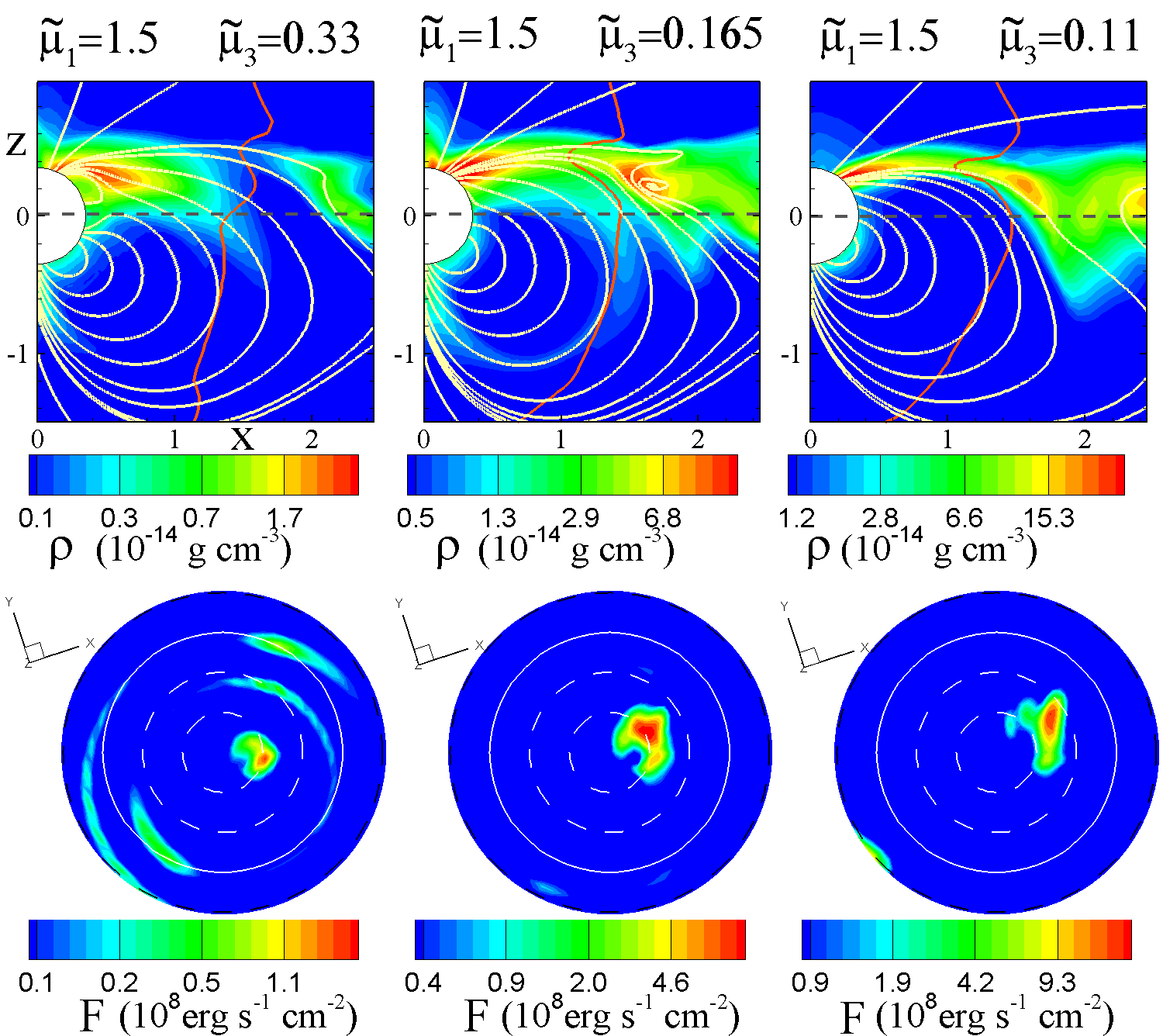}
\caption{\label{dip_4} Test simulations of cases where the ratio
of the dipole to octupole component is twice and thrice as large
compared with the main case (from left to right: models 1, 4 and
5; see also Tab. \ref{tab:mdot}). Top panels show density
distribution (background), selected field lines (yellow lines),
and $\beta=1$ line (red line) in the xz-slices. Bottom panels show
the energy flux distribution on the  star in the plane projection.
Note that in each case we chose such energy level, that the spots
coverage is $f_{tot}\approx 10\%$  and hence it is close to one
observed in V2129 Oph. One can see that the polar spots dominate
in both models 4 and 5.}
\end{center}
\end{figure}

Hence, both simulations and theory point to the fact that a star
with an equatorial dipole magnetic field of 175 G can truncate the
disk at the corotation radius only if the accretion rate is very
small, smaller than that found from observations. Therefore, the
disk should be truncated at smaller distances from the star to
explain the accretion rate.

\subsection{Test simulation runs at larger dipole field component (models 4 and
5)}
\label{sec:mod4and5}

To further resolve this issue, we suggest that the dipole
component is larger than that reconstructed by D07 from
observations. In fact, the July 2009 observations revealed a
dipole component about three times stronger (D10). So, our test
simulations may be relevant to some other moments of time, when
the dipole field is stronger.


We performed two test simulation runs (models 4 and 5) where the
octupolar component is fixed at $B_{3\star}=1.2$ kG, while
 the dipole component is twice and thrice as large:
  $B_{1\star}=700$ G ($\widetilde{\mu}_3/\widetilde{\mu}_1=0.165$) and
 $B_{1\star}=1050$ G  ($\widetilde{\mu}_3/\widetilde{\mu}_1=0.11$).

 Fig. \ref{dip_4} shows the results of these runs versus the main case.
 One can see that in all cases the disk stops at approximately same distance from the star
 (because $\widetilde{\mu}_1=1.5$ in all cases).
 However, the dimensional accretion rate is proportional to $B_{1\star}^2$   and hence it is expected to
 have about
 4 and 9 times higher accretion rate compared with the main case. Exact value also depends on the dimensionless
 parameter $\widetilde{\dot M}$ obtained from simulations.
   Tab. \ref{tab:mdot} (row 12) shows that the accretion rate in these two cases is sufficiently high.
   One can see that even the dipole field of
   $B_{1\star}=700$G is sufficient in disrupting the disk at far distances from the star, and at the same
   time to give the observed accretion rate.
Another advantage of simulations with a higher dipole component is that the main polar spot strongly dominates
which gives a better match with observation of accretion spots in V2129 Oph.

\begin{table*}
\centering
\begin{tabular}{lllllll}
\hline
  Model of V2129 Oph~~  & $\widetilde\mu_1$ & $\widetilde\mu_3$ & $\widetilde{N_f}$ &  $N_f$ (g cm$^2$s$^{-2}$) &  $\tau$ (yr) \\
\hline
model 1.~  dip+oct 3D &  1.5~~ &0.33   & $\pm$0.006  & $\pm 1.7\times 10^{34}$  & $6.5\times 10^8$ & rot. equilibrium \\
model 2.~  dip+oct 3D &  1.0~~ &0.22  & +0.01 &  $+6.2\times 10^{34}$   & $1.7\times 10^8$ & spin-up \\
model 3.~  dip+oct 3D &  0.5~~ &0.11  & +0.015  & $+3.7\times 10^{35}$     &$2.9\times 10^7 $ & spin-up  \\
\hline
model 4.~  dip+oct 3D &  1.5~~ &0.165  & +0.01 &  $+1.1\times 10^{35}$   & $9.7\times 10^7$ & spin-up \\
model 5.~  dip+oct 3D &  1.5~~ &0.11   & $\pm$0.02  & $\pm5.0\times 10^{35}$     &$2.1\times 10^7 $ & rot. equilibrium  \\
\hline
model 6.~  dip 3D     &  1.5~~& 0.00  & $\pm$0.003   & $\pm 8.3\times 10^{33}$ & $1.3\times 10^9$ & rot. equilibrium \\
\hline
\end{tabular}
\caption{The Table shows the dimensionless torque
$\widetilde{N_f}$, dimensional torque $N_f=N_0 \widetilde{N_f}$,
and the time-scale of spinning-up/down $\tau$  for models
1-6.}\label{tab:torque}
\end{table*}

\subsection{Calculation of the torque and spin evolution}
\label{sec:torque}

We also calculate the torque on the star from the
disk-magnetosphere interaction. Matter coming from the disk in
funnel streams can move faster or slower than the star and hence
it creates either a positive or negative azimuthal field on the
star, and correspondingly a positive (spin-up) or negative
(spin-down) magnetic torque, $N_f$. In addition, matter brings
some positive angular momentum which spins the star up, $N_m$. We
calculated both the magnetic, $N_f$, and matter, $N_m$,  torques,
 and found that the torque associated with
the magnetic field  is about 100 times larger than that associated
with matter.  This is typical in accretion through the funnel
streams (see also \citealt{roma02,roma03}). We calculated the
torque for models 1-6 and the results are presented in Table
\ref{tab:torque}. We observed from the simulations that the torque
is small and it wanders around zero taking both positive and
negative values in models 1, 5 and 6 (see Tab. \ref{tab:torque}).
In models 2-4, the torque is slightly positive and the star spins
up. We calculated the dimensional torque $N_f=\widetilde{N_f} N_0$
using corresponding reference values $N_0$ from Tab.
\ref{tab:refval-2}.

Now, we can estimate the time-scale of spin evolution. The angular
velocity of V2129 Oph is (at its period of $P=6.53$ days)
$\Omega=2\pi/P \approx 1.1\times 10^{-5} {\rm s}^{-1}$, its
angular momentum is $J= k M_\star R_\star^2 \Omega =
 8.5\times  10^{50} k ~{\rm g cm}^2/{\rm s}$, where $k<1$.
   Taking $k=0.4$ and different values of $N_f$ for different models from Tab.
 \ref{tab:torque}, we obtain
the characteristic time-scale as $\tau = J/N_f$ (see Tab.
\ref{tab:torque}). One can see that the time-scale in all models
is longer than the  age of V2129 Oph which is estimated to be
$2\times 10^6$ years (see D07). Hence, the torque observed in
simulations is not sufficient to spin down  the star to its
present slow rotation.

If the star has lost it spin in the past due to disk-magnetosphere
interaction, then  we should suggest that in the past the
accretion rate was higher (to ensure rapid spin-down), and the
star's magnetic field was stronger (to ensure disk truncation at a
large distance from the star). In fact it is expected that in the
past, V2129 Oph has been fully convective and its magnetic field
has been amplified due to the dynamo mechanism. The ``propeller"
mechanism can be responsible for the efficient spinning down at
this stage (e.g., \citealt{roma05,ust06}). The stellar winds can
also carry angular momentum out at different stages of star's
evolution (e.g., \citealt{matt05,matt08,cran08,cran09}).

On the other hand, the July 2009 observations by D10 did show that
the magnetic field of the star is stronger than in 2005
observations, and maybe the field of V2129 Oph is higher on an
average than in the June 2005 observations. Note, however, that
the torque is still somewhat small, even when the dipole field is
three times stronger (see Tab. \ref{tab:torque} for model 5).

Note that if a torque at the present epoch is small, and a star is
not in the rotational equilibrium, then the truncation radius can
be at any place (not at the corotation radius).  In such a case
one of the main condition which should be satisfied in comparisons
of models with observations is the similarity of the spot shapes
and positions.

\section{Conclusions and Discussions}
\label{sec:conclusions}

We performed global 3D simulations of disk accretion onto a star
with parameters close those of T Tauri star V2129 Oph.  The
large-scale magnetic field of V2129 Oph is approximated by
slightly tilted dipole and octupole magnetic moments with polar
fields of 0.35 kG and 1.2 kG respectively. Test simulations for
the case of a pure dipole field, smaller magnetospheres and a
larger dipolar field component were performed for comparison.
Below, we summarize our findings and compare our results with the
D07 observations/model:

\smallskip

\noindent {\bf 1.} The field lines connecting the disk and the
star inflate and form magnetic towers above and below the disk.
The potential (vacuum) approximation is still valid inside the
Alfv\'en (magnetospheric) radius, $r_m\lesssim (4-5) R_\star$,
where the magnetic stress dominates over the matter stress. At
larger distances, the magnetic field distribution strongly departs
from the potential one (see Fig. \ref{blinesv}).

\smallskip

\noindent {\bf 2.}  Simulations show that the disk is disrupted by
the dipole component of the field and matter flows towards the
star in two funnel streams. The flow in the stream is redirected
by the octupole component  with part of matter flowing towards the
high-latitude pole, and another part into the low-latitude
octupolar belt. The polar spots are located at high latitudes even
if the disk is truncated close to the star (say, at $r\approx 0.5
R_{cor}$).

\smallskip

\noindent {\bf 3.} In the dipole case the spots are
latitudinally-elongated, they have a typical crescent shapes and
are located  at lower co-latitudes $\theta_c\approx 45^\circ$ (see
also R04). The polar chromospheric spot observed in V2129 Oph
(D07) is much closer in shape and position to the round,
high-latitude polar spot obtained in the dipole+octupole model
than the crescent-shaped spot obtained in the pure dipole model.

\smallskip

\noindent {\bf 4.} If the disk is disrupted at $r_t\approx 6.2
R_\star$, like in our main case (model 1), then the accretion rate
obtained from the simulations is about an order of magnitude
smaller than that obtained from observations. If the disk is
disrupted at $r_t\approx 3.4 R_\star$ (model 3), then the
accretion rate has a good match, but the low-latitude octupolar
belt spots strongly dominate, which is in contrast with
observations. However, if the disk is disrupted at a somewhat
larger distance from the star, $r_t\approx 4.3 R_\star$ (model 2),
then the brightest part of the spot is located in the polar
region, while the octupolar ring spots are much weaker. In this
case the accretion rate is smaller than the observed accretion
rate, but it is within the error bar given by D10 (see Tab.
\ref{tab:mdot}). Hence, model 2 can possibly explain observations
of V2129 Oph. In that case, future, more sensitive observations
should reveal the arcs of the octupolar belt spots.

\smallskip

\noindent{\bf 5.} Both simulations and theoretical estimations
point to the fact that a weak dipole field of $B_{1\star}=350$G
cannot disrupt the disk at large distances from the star, unless
the accretion rate is very low. In a test simulation with a dipole
field twice as strong ($B_{1\star}=700$ G, model 4), the accretion
rate matches the observed accretion rate and the polar spots
dominate. This model explains observations better than the other
models.

\smallskip

\noindent {\bf 6.} We observed that the main (polar) accretion
spot is shifted relative to the main (octupolar)
 magnetic pole (at the phase of 0.15) towards the dipole magnetic pole (see fig. \ref{spot-b}).
 Such a shift may be connected with the fact that matter  flows  toward the dipole magnetic pole and is only partially redirected
 towards the octupolar pole. The dipole and octupole poles differ in phases, and this may lead
to the observed phase difference.
 On the other hand, the funnel stream can be shifted forward/backward relative to the stellar surface
 and the phase of the accretion spot may differ from that of
 the magnetic pole \citep{roma03,roma04a}.

\smallskip

\noindent{\bf 7.} The torque on the star is small and corresponds
to a small accretion rate. The time-scale of spin evolution
($\tau\sim 2\times10^7-10^8$ years) is longer than the estimated
age of V2129 Oph  of $2\times10^6$ years. Hence,
disk-magnetosphere interaction cannot be responsible for the
star's spin evolution at present epoch. Hence, we conclude that
the star probably lost most of its angular momentum during the
early stages of its evolution, for example, at the stage when a
star has been fully convective and had a stronger,
dynamo-generated magnetic field. The ``propeller" mechanism could
be responsible for the rapid spin-down \citep{roma05,usty06}. In
addition, at all stages of evolution angular momentum flows into
stellar winds (e.g. \citealt{matt05,matt08}).

\smallskip

Among the different simulation runs presented here we prefer those
runs where the disk is disrupted at a large distance from the
star, within $(0.7-0.9)R_{cor}$,  in which the star is expected to
be in the rotational equilibrium state (zero torque on average)
(e.g., \citealt{long05}). Such a state is
 probable in slowly rotating CTTSs  \citep{koni91,came93,roma02}. However, the low values of torque
 obtained in our simulations raise the question of whether V2129
 Oph is in the rotational equilibrium state, and hence the position of the inner disk
 is less important for the analysis.

A new set of observations of V2129 Oph  done in July 2009
 have shown that the octupole and dipole components are about
1.5 and 3 times larger compared with the June 2005 observations
discussed in this paper (D10). If the field of V2129 Oph varies
due to the dynamo mechanism, then (at a steady accretion rate) the
disk will move closer to the star during periods when the field is
weaker, and further away when it is stronger. Then, model 2 may
explain the June 2005 observations, and models 4 or 5 (with dipole
moments twice and thrice as large) the July 2009 observations. If
the dipole component is large on an average, say, about (0.7-1)kG,
then the accretion is sufficiently high to explain observations.

\citet{jard08} investigated the properties of the funnel streams
in V2129 Oph  for a fixed magnetic field geometry. They chose a
much higher accretion rate in their model $\dot M\approx
10^{-8}$M$_\odot$/yr, which is a typical accretion rate for mildly
accreting CTTSs (see also D07, \citealt{eisn05}). The density they
obtained in the funnel streams was about a hundred times higher
compared with our model. At this accretion rate, however, the disk
will be truncated very close to the star unless the dipole
magnetic field is even higher than discussed above.

There is an interesting possibility that most of the disk matter
may accrete directly through the equatorial plane due to
 magnetic interchange instabilities \citep{roma08,kulk08,kulk09}, while a smaller amount of matter
climbs the magnetosphere and forms accretion spots. This model can
possibly explain the difference between the low accretion rates
obtained by D10 from spectral lines, and the much higher accretion
rates derived from the disk measurements (e.g., \citealt{eisn05}).

We performed similar 3D MHD simulations of accretion onto another
T Tauri star, BP Tau \citep{long10b} where the dipole and octupole
components of 1.2 kG and 1.6 kG strongly dominate over other
multipoles \citep{dona08}. In that case the dipole component is
much stronger than in V2129 Oph, and there is a better match
between the simulated and observed accretion rates. Both V2129 Oph
and BP Tau are examples of stars with complex fields where the
dipole component determines the disk-magnetosphere interaction,
while octupolar component shapes the spots.

In some T Tauri stars the magnetic field is more complex than in
V2129 Oph, like in  CTTSs CVCha and CRCha \citep{huss09}. Modeling
of such stars requires incorporation of even higher order
multipoles, which can be done in future research.


\section*{Acknowledgments}

Resources supporting this work were provided by the NASA High-End Computing (HEC)
Program through the NASA Advanced Supercomputing (NAS) Division at Ames Research Center and
the NASA Center for Computational Sciences (NCCS) at Goddard Space Flight Center.
 The authors thank A.V. Koldoba and G.V. Ustyugova for the earlier development of
the codes, S.A. Lamzin and R.V.E. Lovelace for discussions, and
the referee for many useful questions and comments. The research
was supported by NSF grant AST0709015, and funds of the Fortner
Endowed Chair at Illinois. The research of MMR was supported by
NASA grant NNX08AH25G and NSF grant AST-0807129.


\end{document}